\documentclass[aps,prl,amsmath,showpacs,preprintnumbers,twocolumn,amssymb,amsmath,superscriptaddress]{revtex4-1}

\usepackage{graphicx}	
\usepackage{dcolumn}	
\usepackage{bm}			
\usepackage{textcomp}
\usepackage[english]{babel}
\usepackage{color}
\usepackage{units}
\definecolor{link:sepia}     {cmyk}{0.00, 0.83, 1.00, 0.70}
\definecolor{link:darkblue}  {cmyk}{1.00, 0.84, 0.10, 0.01}
\definecolor{link:darkgreen} {cmyk}{0.89, 0.34, 1.00, 0.28}
\definecolor{link:black}	 {cmyk}{0.00, 0.00, 0.00, 1.00}
\definecolor{chapred}        {cmyk}{0.29, 0.94, 1.00, 0.36}
\usepackage{hyperref}
\usepackage{multirow}
\usepackage{array}
\newcolumntype{M}[1]{>{\centering\arraybackslash}m{#1}}
\hypersetup{
  bookmarks=false,
  pdffitwindow=true,
  pdfstartview={Fit},
  colorlinks=true,
  linkcolor=link:sepia,		
  citecolor=link:darkblue,	
  urlcolor=link:darkblue,   
}

\begin{document}

\hyphenation{Fesh-bach}
\hyphenation{an-iso-tro-pic}
\hyphenation{as-so-ci-a-tion}

\title{Ultracold Dipolar Molecules Composed of Strongly Magnetic Atoms}
\author{A. Frisch}
\affiliation{Institut f\"{u}r Experimentalphysik, Universit\"{a}t Innsbruck, Technikerstrasse 25, 6020 Innsbruck, Austria}
\affiliation{Institut f\"ur Quantenoptik und Quanteninformation,
 \"Osterreichische Akademie der Wissenschaften, 6020 Innsbruck, Austria}

\author{M. Mark}
\affiliation{Institut f\"{u}r Experimentalphysik, Universit\"{a}t Innsbruck, Technikerstrasse 25, 6020 Innsbruck, Austria}

\author{K. Aikawa}
\altaffiliation{Current address: Department of Physics,	Graduate School of Science and Engineering, Tokyo Institute of Technology, 2-12-1-S5-9, Ookayama, Meguro-ku, Tokyo 152-8551, Japan.}
\affiliation{Institut f\"{u}r Experimentalphysik, Universit\"{a}t Innsbruck, Technikerstrasse 25, 6020 Innsbruck, Austria}

\author{S. Baier}
\affiliation{Institut f\"{u}r Experimentalphysik, Universit\"{a}t Innsbruck, Technikerstrasse 25, 6020 Innsbruck, Austria}

\author{R. Grimm}
\affiliation{Institut f\"{u}r Experimentalphysik, Universit\"{a}t Innsbruck, Technikerstrasse 25, 6020 Innsbruck, Austria}
\affiliation{Institut f\"ur Quantenoptik und Quanteninformation,
 \"Osterreichische Akademie der Wissenschaften, 6020 Innsbruck, Austria}

\author{A. Petrov}
\altaffiliation{Alternative address: NRC ``Kurchatov Institute'' PNPI 188300, Division of Quantum Mechanics, St. Petersburg State University, 198904, Russia.}
\affiliation{Department of Physics, Temple University, Philadelphia, Pennsylvania 19122, USA}

\author{S. Kotochigova}
\affiliation{Department of Physics, Temple University, Philadelphia, Pennsylvania 19122, USA}

\author{G. Qu{\'e}m{\'e}ner}
\affiliation{Laboratoire Aim\'{e} Cotton, CNRS, Universit\'{e} Paris-Sud, ENS Cachan, 91405 Orsay, France}

\author{M. Lepers}
\affiliation{Laboratoire Aim\'{e} Cotton, CNRS, Universit\'{e} Paris-Sud, ENS Cachan, 91405 Orsay, France}

\author{O. Dulieu}
\affiliation{Laboratoire Aim\'{e} Cotton, CNRS, Universit\'{e} Paris-Sud, ENS Cachan, 91405 Orsay, France}

\author{F. Ferlaino}
\affiliation{Institut f\"{u}r Experimentalphysik, Universit\"{a}t Innsbruck, Technikerstrasse 25, 6020 Innsbruck, Austria}
\affiliation{Institut f\"ur Quantenoptik und Quanteninformation,
 \"Osterreichische Akademie der Wissenschaften, 6020 Innsbruck, Austria}

\date{\today}

\begin{abstract}
	In a combined experimental and theoretical effort, we demonstrate a novel type of dipolar system made of ultracold bosonic dipolar molecules with large magnetic dipole moments. Our dipolar molecules are formed in weakly bound Feshbach molecular states from a sample of strongly magnetic bosonic erbium atoms. We show that the ultracold magnetic molecules can carry very large dipole moments and we demonstrate how to create and characterize them, and how to change their orientation. Finally, we confirm that the relaxation rates of molecules in a quasi-two dimensional geometry can be reduced by using the anisotropy of the dipole-dipole interaction and that this reduction follows a universal dipolar behavior.
\end{abstract}

\pacs{34.50.-s, 31.10.+z, 34.50.Cx}



\maketitle
Ultracold dipolar particles are at the heart of very intense research activities, which aim to study the effect of interactions that are anisotropic and long range~\cite{Baranov2012cmt,Lahaye2009tpo}. Dipolar quantum phenomena require ultracold gases and a strong dipole-dipole interaction (DDI). So far, strongly dipolar gases have been obtained using either atoms with a large magnetic dipole moment or ground-state polar molecules with an electric dipole moment\,\cite{Lahaye2009tpo}. With both systems, many fascinating many-body quantum effects have been observed and studied, such as the $d$-wave collapse of a dipolar Bose-Einstein condensate~\cite{Lahaye2007sde,Aikawa2012bec}, the deformation of the Fermi sphere~\cite{Aikawa2014oof}, and the spin-exchange phenomena~\cite{Yan2013ood,DePaz2013nqm}.

Here, we introduce a novel kind of strongly dipolar particles. These are weakly bound dipolar molecules produced from a pair of atoms with large magnetic dipole moments, such as erbium (Er). The central idea is that these molecules can possess a very large magnetic moment $\mu$ up to twice that of atoms (e.\,g.\,$14$ Bohr magneton, $\mu_B$, for Er$_2$) and have twice the mass of the atoms. As a consequence, the degree of ``dipolarity'' of the magnetic molecules is much larger than the one of atoms. This can be quantified in terms of the dipolar length $a_d= m \mu_0 \mu^2/(4 \pi \hbar^2)$~\cite{Baranov2012cmt}, which solely depends on the molecular mass $m$ and on $\mu$; $\hbar$ is the Planck constant divided by $2\pi$. To give an example, Er$_2$ with $\mu=14\,\mu_B$ has a $a_d$ of about $1600\,a_0$, which largely exceeds the typical values of the $s$-wave scattering length.  Here, $a_0$ is the Bohr radius. Moreover, in contrast to ground-state heteronuclear molecules, the dipole moment of the magnetic molecules does not vanish at zero external (magnetic) field, opening the intriguing possibility of investigating the physics of unpolarized dipoles.

In a joined experimental and theoretical effort, we study the key aspects of ultracold dipolar Er$_2$ molecules, including the association process, the molecular energy spectrum, the magnetic dipole moments, and the scattering properties in both three- (3D) and quasi two-dimensional (Q2D) geometries.

Erbium belongs to the class of strongly magnetic lanthanides, which are currently attracting great attention in the field of ultracold quantum gases~\cite{Lu2011sdb,Sukachev2010mot,Aikawa2012bec,Miao2014mot}. 
Indeed, these species exhibit unique interactions. 
Beside the long-range magnetic DDI, these species have both an isotropic and an anisotropic contribution in the short range van der Waals (vdW) potential. The latter results from the large non-zero orbital momentum quantum number of the atoms~\cite{Petrov2012aif,Lepers2014aot}.  This manifold leads to an extraordinary rich molecular spectrum, reflecting itself in a likewise dense spectra of Feshbach resonances as demonstrated in recent scattering experiments~\cite{Aikawa2012bec,Frisch2014qci,Baumann2014ool}. Each resonance position marks an avoided crossing between the atomic scattering threshold and a molecular bound state, which can be used to associate molecules from atom pairs~\cite{Chin2010fri}. 

We create and probe Er$_2$ dipolar molecules by using standard magneto-association and imaging techniques~\cite{Chin2010fri}. Details of the production schemes are described in the Supplemental Material~\cite{SupplementMat}. In brief, we begin with an ultracold sample of $^{168}$Er atoms in an optical dipole trap (ODT) in a crossed-beam configuration. The atoms are spin-polarized into the lowest Zeeman sublevel ($j=6$, $m_{j}=-6$). Here, $j$ is the atomic electronic angular momentum quantum number and $m_j$ is its projection on the quantization axis along the magnetic field. To associate Er$_2$ molecules, we ramp the magnetic field across one of the low-field Feshbach resonances observed in Er~\cite{Aikawa2012bec,Frisch2014qci}. We experimentally optimize the ramping parameters, such as the ramp speed and the magnetic-field sweep interval, by maximizing the conversion efficiency. In our experiment we typically achieve a conversion efficiency of $\unit[15]{\%}$, which is a common value for boson-composed Feshbach molecules~\cite{Chin2010fri}. To obtain a pure molecular sample, we remove all the remaining atoms from the ODT by applying a resonant laser pulse. Our final molecular sample contains about $2\times10^4$ Er$_2$ Feshbach molecules at a temperature of $\unit[300]{nK}$ and at a density of about $\unit[8\times10^{11}]{cm^{-3}}$~\cite{SupplementMat}.

\begin{figure}
	\centering
	\includegraphics[width=1\linewidth]{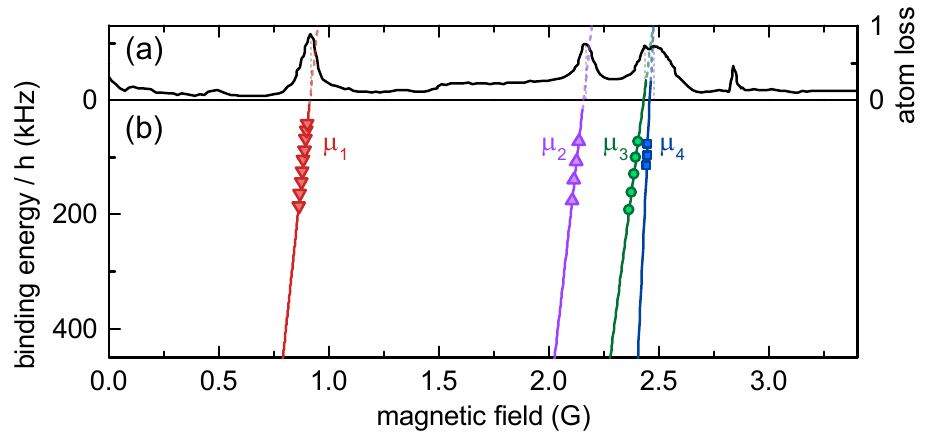}
	\caption{(color online). Er$_2$ weakly bound molecules. (a) Atom-loss spectrum \cite{Aikawa2012bec} from $0$ to $\unit[3]{G}$ and (b) near-threshold binding energy of the corresponding molecular states. The solid lines are fits to the experimental data and extrapolated to larger $E_{b}$ up to $ h \times \unit[500]{kHz}$. The error bars are smaller than the symbols.}
	\label{fig:spectroscopy}
\end{figure}

A central question regards the magnitude of the dipole moment owned by the magnetic molecules. We experimentally determine $\mu$ by using magnetic-field modulation spectroscopy, a technique which was successfully applied to alkali atoms~\cite{Claussen2003vhp,Thompson2005ump,Mark2007sou}. With this method, we measure the molecular binding energy $E_{b}$ near the atomic threshold as a function of the magnetic field $B$. The binding energy is related to the differential magnetic moment of the molecules with respect to the atom-pair magnetic moment $2a$. Here, $\mu_a=-g m_j \mu_B=6.98\,\mu_B$ in the case of Er, where $g = 1.16$ is the Er atomic Land\'e factor. We thus extract $\mu$ by using the relation $\mu=2\mu_a-|dE_{b}(B)/dB|$. Our spectroscopic measurement begins with an ultracold atomic sample near a Feshbach resonance. We then add a small sinusoidal modulation to the bias magnetic field for $\unit[400]{ms}$. The modulation frequency is varied at each experimental run. When it matches $E_{b}/h$, prominent atom losses appear because of molecule formation. We trace the near-threshold molecular spectrum by repeating the measurement for various magnetic-field values. Figure\,\ref{fig:spectroscopy} shows the Er$_2$ molecular spectrum in a magnetic field range up to $\unit[3]{G}$. In our range of investigation, we identify four molecular energy levels, which near treshold exhibit a linear dependence on $B$. For each state, we obtain a different $\mu$ value, ranging from $8$ to $12\,\mu_B$ \footnote{For future experiments, it will be very interesting to access molecular levels parallel to the atomic threshold for which $\mu=2\mu_a$~\cite{Mark2007sou}. 
}, as listed in Table~\ref{tab:magneticmoments}.

\begin{table}
	\caption{Experimental and theoretical magnetic moments of four molecular states near the atomic threshold, Feshbach-resonance positions $B_{\mathrm{FR}}$, dipolar lengths, outer turning points $R^*$, and dominant quantum numbers $\ell$, $J$, and $M$. For convenience, the molecular states are labeled as $\mu_i$ with $i=1,\dots 4$. The specified uncertainties correspond to the $1\sigma$ statistical errors.}
	\begin{ruledtabular}
		\begin{tabular}{M{3mm} M{7mm}|M{11mm} M{8mm}|M{12mm} M{6mm} l}
			& $B_{\mathrm{FR}}$  & \multicolumn{2}{c|}{$\mu/\mu_B$} &$a_d$ & $R^*$ &  $|\ell$,$J$,$M\rangle$ \\
			& (G) & Exp. & Theo.& ($a_0$) & ($a_0$) &     \\
			\hline 
			$\mu_1$ & $0.91$ & $11.30(7)$ & $11.20$ & $1041(13)$ & $72.0$ & $|4$,$12$, $-12$/$-10$/$-9\rangle $ \\
			$\mu_2$ & $2.16$ & $11.51(4)$ & $11.46$ & $1080(8)$  & $71.0$ & $|4$,$10$, $-10\rangle$ \\
			$\mu_3$ & $2.44$ & $11.84(2)$ & $11.75$ & $1143(4)$  & $86.0$ & $|2$,$12$, $-10\rangle$ \\
			$\mu_4$ & $2.47$ & $7.96(3)$  & $7.92$  & $517(4)$   & $57.0$ & $|6$,$10$, $-7$/$-6\rangle$ \\
		\end{tabular} 
	\end{ruledtabular}
	\label{tab:magneticmoments}
\end{table}

For alkali-metal atoms, which possess much simpler interaction properties than lanthanides, theoretical approaches based on coupled-channel calculations have been extremely successful in assigning the quantum numbers of the molecular energy levels and reproducing molecular spectra~\cite{Chin2010fri}. However, a straightforward extension of these methods to the lanthanide case is out of reach because of their complex scattering physics involving highly anisotropic interactions and many partial waves~\cite{Frisch2014qci}. Inspired by work on alkali-metal collisions~\cite{Moerdijk1995riu,Stan2004oof,Wille2008eau,Tiecke2010abs}, we develop a new theoretical approach to identify the molecular quantum numbers, based on approximate adiabatic potentials and on the experimentally measured $\mu$ as input parameters. Our scattering model is detailed in the Supplemental Material~\cite{SupplementMat}, whereas we here summarize the central ideas of our approach.

We first solve the eigenvalue problem of the full atom-atom interaction potential operator~\cite{SupplementMat}, whose eigenvalues are the adiabatic potentials $U_n(R;B)$. The corresponding eigenfunctions read as $|n;R\rangle=\sum_{i} c_{n,i}(R)|i\rangle$, where $n=1,2,\dots$ and $c_{n,i}(R)$ are $R$-dependent coefficients. The molecular state $|i\rangle$ is uniquely determined by the set of angular momentum quantum numbers ($\ell$, $J$, $M$), where $\ell$ is the molecular orbital quantum number, $\vec J=\vec j_1 +\vec j_2$ the total atomic angular momentum,  and $M$ its projection on the internuclear axis.

To derive the corresponding ``adiabatic'' molecular magnetic moments, we calculate $\mu^{\rm calc}\approx-dU_n(R;B)/dB$ at the position of the outer classical turning point $R=R^*$. This choice is justified by the fact that most of the vibrational wavefunction is localized around $R^*$.

\begin{figure}
	\centering
	\includegraphics[width=0.9\columnwidth]{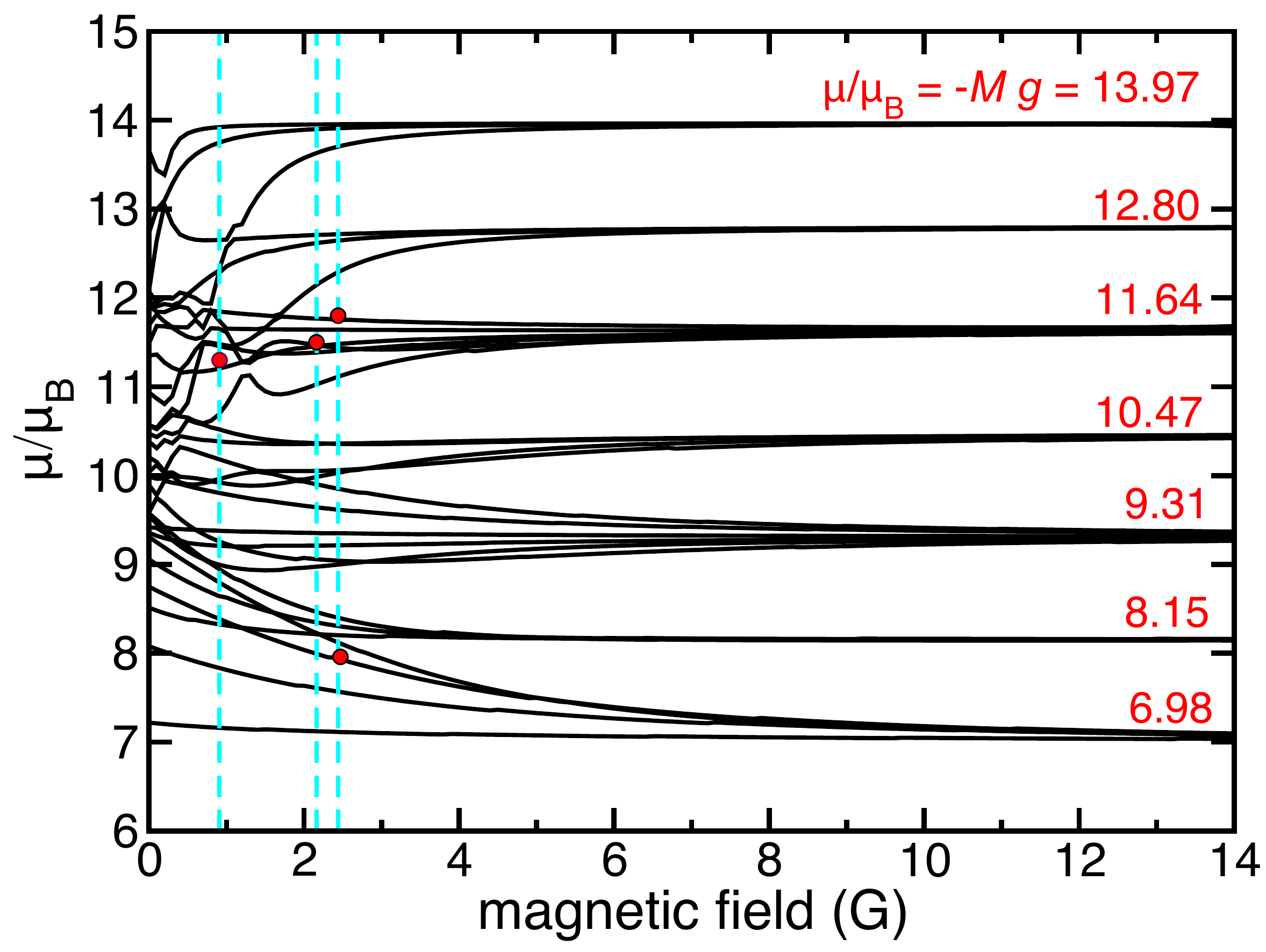}
	\caption{(color online). Adiabatic magnetic moment as a function of magnetic field strength evaluated at the entrance channel energy. Each curve corresponds to the adiabatic magnetic moment of one adiabatic potential $U_n(R;B)$. The magnetic moments in the asymptotic limit of large $B$ are given. The dashed vertical lines correspond to the field strength where we have observed Feshbach resonances. The red-filled circles represent the experimentally measured magnetic moments at these resonance locations.}
	\label{fig:MagMom}
\end{figure}

From the Hellmann-Feynman theorem it then follows that $\mu^{\rm calc}=-g\mu_B \sum_{i}M_i|c_{i}(R^*)|^2$.
Finally, we assume that for each Feshbach resonance a vibrational state is on resonance and we find the adiabatic potential that has a magnetic moment closest to the measured one within $1\,\%$.
Once the best match is identified, the corresponding $|n;R\rangle$ sets the molecular state $|i\rangle$, characterized by $\ell$, $J$, and $M$, with the largest, dominant contribution. In our range of investigation we observe $d$-, $g$-, and $i$-wave molecular states; see Table~\ref{tab:magneticmoments}. These states show several dominant $M$ contributions. This fact is unusual and reflects the dominant role of the DDI, which couples several adiabatic potentials and $M$ components. As shown in Fig.~\ref{fig:MagMom}, this mixing effect is particularly dominant below $\unit[10]{G}$, where the DDI at $R^*$ is larger than the Zeeman interaction. Above $\unit[10]{G}$, we predict $\mu$ to be equal to integer multiples of $g\mu_B$~\cite{SupplementMat}. 

\begin{figure}
	\centering
	\includegraphics[width=1\columnwidth]{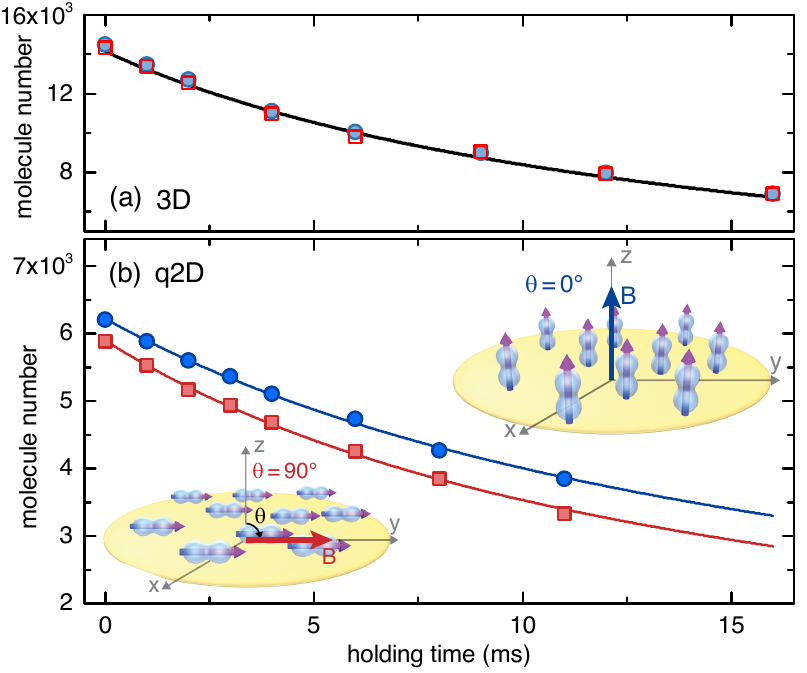}
	\caption[Fig3]{(color online). Typical time evolution of the number of molecules for $\theta=90^{\circ}$ (squares) and $\theta=0^{\circ}$ (circles) in a 3D (a) and in a Q2D trap (b). The data refer to molecules in the state $\mu_1$ for the 3D case (a) and molecules in the state $\mu_2$ in Q2D (b). The insets in (b) show an illustration of molecules in pancake-shaped traps with out-of-plane (right) and  in-plane (left) orientations. The solid lines are two-body decay fits to the data. The error bars for (a) and (b) are smaller than the data points and are not shown. The data points in (a) are obtained by averaging $5$ independent measurements and in (b) about $50$ measurements have been averaged.}
	\label{fig:decaycurves}
\end{figure}

As summarized in Table~\ref{tab:magneticmoments}, we find very good overall agreement between the measured and the calculated molecular magnetic moments. For the largest observed $\mu$, we calculate a corresponding dipolar length $a_d\approx \unit[1150]{a_0}$. This value exceeds the typical range of the vdW potentials, setting the DDI as the dominant interaction in the system. Remarkably, $a_d$ for Er$_2$ is comparable to the one realized with ground-state KRb molecules\,\cite{Ni2010dco}, which are an extensively investigated case serving as a benchmark dipolar system. 

Following the methods introduced for KRb \cite{Quemener2010efs,Micheli2010urf,Miranda2011ctq}, we test the dipolar character of Er$_2$ by performing scattering experiments in a three- (3D) and in a quasi two-dimensional  (Q2D) optical dipole trap. We control the DDI between molecules by tuning the dipole orientation, which is controlled by changing the direction of the magnetic field and is represented by the angle $\theta$ between the magnetic field axis and the gravity axis. Our experiment begins with the atomic sample trapped either in a 3D or in a Q2D ODT. The Q2D trap is created by superimposing a vertically oriented, one-dimensional optical lattice~\cite{SupplementMat}. After the magneto-association and the removal of the remaining atoms, we probe the number of molecules as a function of the holding time in the ODT. We perform measurements for the molecular states $\mu_1$, $\mu_2$, and $\mu_4$ \footnote{The state $\mu_3$ is difficult to access because of its proximity with $\mu_4$.}. For each of these states, we measure the collisional stability of the sample for both in-plane ($\theta=90^{\circ}$) and out-of-plane ($\theta=0^{\circ}$) dipole orientation, and extract the corresponding relaxation rate coefficients, $\beta_\perp$ and $\beta_\|$, using a standard two-body rate equation \cite{Ferlaino2010cou}. 

Figure~\ref{fig:decaycurves} shows typical molecular decay curves in (a) 3D and in (b) Q2D. In 3D, we confirm that the inelastic decay does not depend on $\theta$. We obtain $\beta_{3\mathrm{D}}=\unit[1.3(2) \times 10^{-10}]{cm^3/s}$. This is a typical value for boson-composed Feshbach molecules, which undergo a rapid vibrational quenching into lower-lying molecular states, as demonstrated with alkali atoms \cite{Ferlaino2010cou}. Contrary, in Q2D the decay rates clearly depend on the dipole orientation.  For each investigated molecular state, $\beta_\perp$ is larger than $\beta_\|$. We find a reduction of losses of up to $30\,\%$ for out-of-plane orientation, for which the DDI is predominantly repulsive. The ratio $\frac{\beta_\perp(T)}{\beta_\|(T)}$ increases with increasing $\mu$; see Table~\ref{tab:lossrate}.  We note that stronger suppression of losses can be obtained using a tighter two-dimensional confinement \cite{Quemener2010efs}, which is presently not reachable with our experimental parameters.

The reduction of losses in Q2D draws a natural analogy with the observations obtained with KRb molecules \cite{Miranda2011ctq}.   From a comparative analysis between Er$_2$ and KRb, one can unveil universal behaviour attributed to the DDI, for systems being different in nature, but sharing a similar degree of dipolarity. We thus theoretically study the scattering behavior of Er$_2$ using a  theoretical approach similar to the one successfully applied to KRb. Our formalism, which accounts for the DDI and the isotropic vdW interaction, is described in Refs.\,\cite{SupplementMat, Quemener2015dou}.

We compute the ${\mathrm{Er}_2+\mathrm{Er}_2}$ loss rate coefficients $\beta(T)$ in 3D and in Q2D for given values of $\mu$, $\theta$, and $T$. By averaging over a 3D and a 2D Maxwell-Boltzmann distribution, we obtain the thermalized loss rate coefficients $\beta(T)$ in 3D and in Q2D, respectively. In 3D, we find a rate coefficient of $\unit[1.0 \times 10^{-10}]{cm^3/s}$ at $T=\unit[300]{nK}$, which is close to the experimental value~\footnote{Note that this value is two orders of magnitude larger than that of $p$-wave dominated fermionic KRb molecules~\cite{Ni2010dco}.}. In Q2D, our calculations show that the collision dynamics at long range, and thus the value of $\beta$, depends on the dipole orientation and monotonically increases with $\mu$. As in the experiments, our calculations show that collisions for in-plane orientation ($\beta_\perp$) lead to larger molecular losses than for out-of-plane orientation ($\beta_\|$). In Table~\ref{tab:lossrate}, we compare theory and experiment. The absolute values of $\beta$ agree within a factor of two. This difference is well explained by the fact that our model does not include details of the short-range physics, being the Er$_4$ potential energy surfaces currently unknown~\cite{SupplementMat}. 

\begin{table}
	\caption{Experimental and theoretical loss rate coefficients $\beta$ for $T=\unit[400]{nK}$ and for various $\mu$ and $\theta$ at $B=\unit[200]{mG}$. Uncertainties of $\beta$ are statistical from fitting and systematic due to number density uncertainty. For the slightly different values of $\mu$ compared to Table~\ref{tab:magneticmoments} and the error discussion see Supplemental Material~\cite{SupplementMat}.}
	\begin{ruledtabular}
		\begin{tabular}{M{5mm}|M{10mm}|M{20mm} M{9mm} |M{20mm} M{9mm}}
			& $\mu/\mu_B$ & \multicolumn{2}{c|}{$\beta_\perp$~($\unit[10^{-6}]{cm^2/s}$)} & \multicolumn{2}{c}{$\beta_\|$~($\unit[10^{-6}]{cm^2/s}$)} \\
			&  & Exp. & Theo. & Exp. & Theo.  \\
			\hline
			{$\mu_4$} & $8.7(6)$  & $12.5\pm0.3\pm3.3$ & $6.00 $  & $10.6\pm0.3\pm2.8$  & $4.79 $ \\
			{$\mu_1$} & $10.9(5)$ & $9.5\pm0.2\pm2.5$  & $6.81 $  & $7.3\pm0.1\pm2.1$   & $5.07 $ \\
			{$\mu_2$} & $11.7(3)$ & $11.3\pm0.2\pm2.9$ & $7.12 $  & $8.6\pm0.2\pm2.3$   & $5.13 $
		\end{tabular}
	\end{ruledtabular}
	\label{tab:lossrate}
\end{table}

\begin{figure}
	\centering
	\includegraphics[width=1\columnwidth]{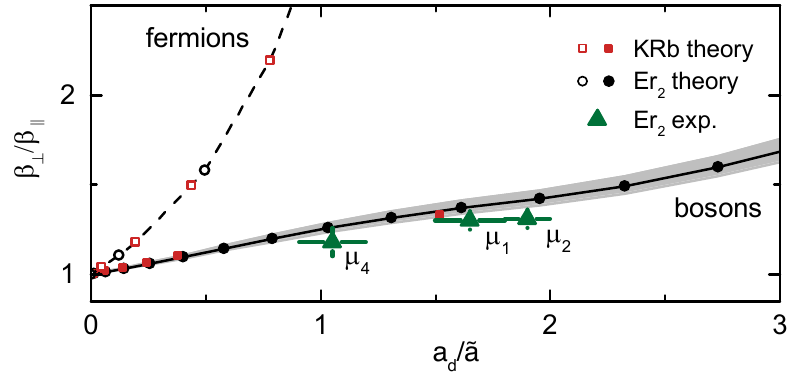}
	\caption{(color online). Universal loss rate ratio $\beta_\perp / \beta_\|$ as a function of $a_d/{\tilde a}$ for Er$_2$ (circles) and KRb (squares) for a fixed value $ a_{\mathrm{dB}} / a_{\mathrm{ho}} = 4.85$ corresponding to $T=\unit[400(40)]{nK}$ and $\nu_z = \unit[31.2(1)]{kHz}$~\cite{SupplementMat}. The gray shaded area is due to the uncertainty of $T$. Here, $\tilde a=a_{\mathrm{ho}}$ for bosonic molecules (filled symbols) and $\tilde a=a_{\mathrm{vdW}}$ for fermionic molecules (open symbols). The calculated loss rate ratios of Er$_2$ are compared with the experimental data for states $\mu_1$, $\mu_2$, and $\mu_4$ (triangles).}
	\label{fig:ratiolossrate}
\end{figure}

Remarkably, the experimental and calculated ratios $\frac{\beta_\perp(T)}{\beta_\|(T)}$ agree very well with each other; see Fig.\,\ref{fig:ratiolossrate}. This suggests that $\frac{\beta_\perp(T)}{\beta_\|(T)}$ for Er$_2$ Feshbach molecules is determined by the DDI and not by the short-range physics, and that it can be correctly described using a point-like-dipole formalism~\cite{SupplementMat}. Figure \ref{fig:ratiolossrate} shows the comparative analysis between bosonic $^{41}$K$^{87}$Rb and $^{168}$Er$_2$, and fermionic $^{40}$K$^{87}$Rb and $^{167}$Er$^{168}$Er based on our numerical calculations. Independent of the nature of the magnetic or electric dipolar system, we find universal curves as a function of $a_d/\tilde{a}$: one for bosons with $\tilde{a}=a_{\rm ho}$ and one for fermions when $\tilde{a}=a_{\rm vdW}$. Here, $a_{\mathrm{ho}}$ is the harmonic oscillator length and $a_{\mathrm{vdW}}=(2mC_6/\hbar^2)^{1/4}$ is the vdW length with $C_6$ the vdW coefficient. The faster increase of $\frac{\beta_\perp}{\beta_\|}$ for fermions with respect to bosons is due to the statistical fermionic suppression of $\beta_\|$ in Q2D that does not occur for bosons as explained in Ref.\,\cite{Quemener2011dou}.

The universal behavior of ultracold dipolar scattering has been previously pointed out in Ref.\,\cite{Quemener2011uiu}. In the Wigner regime, we derive simple universal scaling laws for dipolar bosonic and fermionic molecules~\cite{SupplementMat,Quemener2011uiu}. For bosons with $a_d, a_{\mathrm{ho}} > a_{\mathrm{vdW}} $, which is the case of    our Er$_2$ molecules, we find $\frac{\beta_\perp(T)}{\beta_\|(T)}\sim\big(\frac{a_{\mathrm{dB}}}{a_{\mathrm{ho}}}\big)^4 \, \frac{a_d}{a_{\mathrm{ho}}}
\, \mathrm{exp}\Big[2\big(\frac{a_d}{a_{\mathrm{ho}}}\big)^{2/5}\Big]$. For fermions with $a_d, a_{\mathrm{vdW}} < a_{\mathrm{ho}} $, $\frac{\beta_\perp}{\beta_\|}\sim\big(\frac{a_d}{a_{\mathrm{vdW}}}\big)^3$. Here, $a_\mathrm{dB} = h/\sqrt{2 \pi m k_{\mathrm{B}}T}$ is the thermal de Broglie wavelength.

To conclude, our work reports on the study of strongly dipolar molecules created by pairing ultracold atoms with large magnetic dipole moments. We anticipate that our scheme can be generalized to other magnetic lanthanide species and  has the potential to open regimes of investigations, which have been unaccessible so far. First, the extraordinarily dense and rich molecular energy spectrum of Er opens the exciting prospect of cruising through molecular states of different magnetic moments or even creating molecular-state mixtures with dipole imbalance \cite{Mark2007siw,Mark2007sou,Lang2008ctm}. Second, in contrast to electric polar molecules where the electric dipole moment is zero in the absence of a polarizing electric field, magnetic dipolar molecules have a permanent dipole moment allowing to study the physics of unpolarized dipoles. In addition, strongly magnetic Feshbach molecules offer a novel case of study for scattering physics. These molecules are in fact diffuse in space with a typical size on the order of the vdW length. This novel situation can also have interesting consequences and trigger the development of extended scattering models, which account for multi-polar effects and truly four-body contributions when the molecule size becomes comparable to $a_d$~\cite{Quemener2015fbl}. Finally, a very promising development will be to create fermionic Er$_2$ dipolar molecules where vibrational quenching processes are intrinsically suppressed because of the Pauli exclusion principle~\cite{Greiner2003eoa,Jochim2003bec}.

\begin{acknowledgments}
	The Innsbruck team acknowledges support by the Austrian Ministry of Science and Research (BMWF) and the Austrian Science Fund (FWF) through a START Grant under Project No. Y479-N20 and by the European Research Council under Project No. 259435. Research at Temple University is supported by an AFOSR grant No. FA9550-14-1-0321 and NSF grant No. PHY-1308573. G.Q., M.L. and O.D. acknowledge the financial support of ``Projet Attractivit\'e 2014'' from Universit\'e Paris-Sud and the COPOMOL project (\#ANR-13-IS04-0004) from Agence Nationale de la Recherche.
\end{acknowledgments}


\begin{thebibliography}{42}%
	\makeatletter
	\providecommand \@ifxundefined [1]{%
		\@ifx{#1\undefined}
	}%
	\providecommand \@ifnum [1]{%
		\ifnum #1\expandafter \@firstoftwo
		\else \expandafter \@secondoftwo
		\fi
	}%
	\providecommand \@ifx [1]{%
		\ifx #1\expandafter \@firstoftwo
		\else \expandafter \@secondoftwo
		\fi
	}%
	\providecommand \natexlab [1]{#1}%
	\providecommand \enquote  [1]{``#1''}%
	\providecommand \bibnamefont  [1]{#1}%
	\providecommand \bibfnamefont [1]{#1}%
	\providecommand \citenamefont [1]{#1}%
	\providecommand \href@noop [0]{\@secondoftwo}%
	\providecommand \href [0]{\begingroup \@sanitize@url \@href}%
	\providecommand \@href[1]{\@@startlink{#1}\@@href}%
	\providecommand \@@href[1]{\endgroup#1\@@endlink}%
	\providecommand \@sanitize@url [0]{\catcode `\\12\catcode `\$12\catcode
		`\&12\catcode `\#12\catcode `\^12\catcode `\_12\catcode `\%12\relax}%
	\providecommand \@@startlink[1]{}%
	\providecommand \@@endlink[0]{}%
	\providecommand \url  [0]{\begingroup\@sanitize@url \@url }%
	\providecommand \@url [1]{\endgroup\@href {#1}{\urlprefix }}%
	\providecommand \urlprefix  [0]{URL }%
	\providecommand \Eprint [0]{\href }%
	\providecommand \doibase [0]{http://dx.doi.org/}%
	\providecommand \selectlanguage [0]{\@gobble}%
	\providecommand \bibinfo  [0]{\@secondoftwo}%
	\providecommand \bibfield  [0]{\@secondoftwo}%
	\providecommand \translation [1]{[#1]}%
	\providecommand \BibitemOpen [0]{}%
	\providecommand \bibitemStop [0]{}%
	\providecommand \bibitemNoStop [0]{.\EOS\space}%
	\providecommand \EOS [0]{\spacefactor3000\relax}%
	\providecommand \BibitemShut  [1]{\csname bibitem#1\endcsname}%
	\let\auto@bib@innerbib\@empty
	\bibitem [{\citenamefont {Baranov}\ \emph {et~al.}(2012)\citenamefont
		{Baranov}, \citenamefont {Dalmonte}, \citenamefont {Pupillo},\ and\
		\citenamefont {Zoller}}]{Baranov2012cmt}%
	\BibitemOpen
	\bibfield  {author} {\bibinfo {author} {\bibfnamefont {M.~A.}\ \bibnamefont
			{Baranov}}, \bibinfo {author} {\bibfnamefont {M.}~\bibnamefont {Dalmonte}},
		\bibinfo {author} {\bibfnamefont {G.}~\bibnamefont {Pupillo}}, \ and\
		\bibinfo {author} {\bibfnamefont {P.}~\bibnamefont {Zoller}},\ }\href
	{\doibase 10.1021/cr2003568} {\bibfield  {journal} {\bibinfo  {journal}
			{Chem. Rev.}\ }\textbf {\bibinfo {volume} {112}},\ \bibinfo {pages} {5012}
		(\bibinfo {year} {2012})}\BibitemShut {NoStop}%
	\bibitem [{\citenamefont {Lahaye}\ \emph {et~al.}(2009)\citenamefont {Lahaye},
		\citenamefont {Menotti}, \citenamefont {Santos}, \citenamefont {Lewenstein},\
		and\ \citenamefont {Pfau}}]{Lahaye2009tpo}%
	\BibitemOpen
	\bibfield  {author} {\bibinfo {author} {\bibfnamefont {T.}~\bibnamefont
			{Lahaye}}, \bibinfo {author} {\bibfnamefont {C.}~\bibnamefont {Menotti}},
		\bibinfo {author} {\bibfnamefont {L.}~\bibnamefont {Santos}}, \bibinfo
		{author} {\bibfnamefont {M.}~\bibnamefont {Lewenstein}}, \ and\ \bibinfo
		{author} {\bibfnamefont {T.}~\bibnamefont {Pfau}},\ }\href
	{http://stacks.iop.org/0034-4885/72/i=12/a=126401} {\bibfield  {journal}
		{\bibinfo  {journal} {Rep. Prog. Phys.}\ }\textbf {\bibinfo {volume} {72}},\
		\bibinfo {pages} {126401} (\bibinfo {year} {2009})}\BibitemShut {NoStop}%
	\bibitem [{\citenamefont {Lahaye}\ \emph {et~al.}(2007)\citenamefont {Lahaye},
		\citenamefont {Koch}, \citenamefont {Fr\"ohlich}, \citenamefont {Fattori},
		\citenamefont {Metz}, \citenamefont {Griesmaier}, \citenamefont
		{Giovanazzi},\ and\ \citenamefont {Pfau}}]{Lahaye2007sde}%
	\BibitemOpen
	\bibfield  {author} {\bibinfo {author} {\bibfnamefont {T.}~\bibnamefont
			{Lahaye}}, \bibinfo {author} {\bibfnamefont {T.}~\bibnamefont {Koch}},
		\bibinfo {author} {\bibfnamefont {B.}~\bibnamefont {Fr\"ohlich}}, \bibinfo
		{author} {\bibfnamefont {M.}~\bibnamefont {Fattori}}, \bibinfo {author}
		{\bibfnamefont {J.}~\bibnamefont {Metz}}, \bibinfo {author} {\bibfnamefont
			{A.}~\bibnamefont {Griesmaier}}, \bibinfo {author} {\bibfnamefont
			{S.}~\bibnamefont {Giovanazzi}}, \ and\ \bibinfo {author} {\bibfnamefont
			{T.}~\bibnamefont {Pfau}},\ }\href {\doibase 10.1038/nature06036} {\bibfield
		{journal} {\bibinfo  {journal} {Nature}\ }\textbf {\bibinfo {volume} {448}},\
		\bibinfo {pages} {672} (\bibinfo {year} {2007})}\BibitemShut {NoStop}%
	\bibitem [{\citenamefont {Aikawa}\ \emph {et~al.}(2012)\citenamefont {Aikawa},
		\citenamefont {Frisch}, \citenamefont {Mark}, \citenamefont {Baier},
		\citenamefont {Rietzler}, \citenamefont {Grimm},\ and\ \citenamefont
		{Ferlaino}}]{Aikawa2012bec}%
	\BibitemOpen
	\bibfield  {author} {\bibinfo {author} {\bibfnamefont {K.}~\bibnamefont
			{Aikawa}}, \bibinfo {author} {\bibfnamefont {A.}~\bibnamefont {Frisch}},
		\bibinfo {author} {\bibfnamefont {M.}~\bibnamefont {Mark}}, \bibinfo {author}
		{\bibfnamefont {S.}~\bibnamefont {Baier}}, \bibinfo {author} {\bibfnamefont
			{A.}~\bibnamefont {Rietzler}}, \bibinfo {author} {\bibfnamefont
			{R.}~\bibnamefont {Grimm}}, \ and\ \bibinfo {author} {\bibfnamefont
			{F.}~\bibnamefont {Ferlaino}},\ }\href {\doibase
		10.1103/PhysRevLett.108.210401} {\bibfield  {journal} {\bibinfo  {journal}
			{Phys. Rev. Lett.}\ }\textbf {\bibinfo {volume} {108}},\ \bibinfo {pages}
		{210401} (\bibinfo {year} {2012})}\BibitemShut {NoStop}%
	\bibitem [{\citenamefont {Aikawa}\ \emph {et~al.}(2014)\citenamefont {Aikawa},
		\citenamefont {Baier}, \citenamefont {Frisch}, \citenamefont {Mark},
		\citenamefont {Ravensbergen},\ and\ \citenamefont
		{Ferlaino}}]{Aikawa2014oof}%
	\BibitemOpen
	\bibfield  {author} {\bibinfo {author} {\bibfnamefont {K.}~\bibnamefont
			{Aikawa}}, \bibinfo {author} {\bibfnamefont {S.}~\bibnamefont {Baier}},
		\bibinfo {author} {\bibfnamefont {A.}~\bibnamefont {Frisch}}, \bibinfo
		{author} {\bibfnamefont {M.}~\bibnamefont {Mark}}, \bibinfo {author}
		{\bibfnamefont {C.}~\bibnamefont {Ravensbergen}}, \ and\ \bibinfo {author}
		{\bibfnamefont {F.}~\bibnamefont {Ferlaino}},\ }\href {\doibase
		10.1126/science.1255259} {\bibfield  {journal} {\bibinfo  {journal}
			{Science}\ }\textbf {\bibinfo {volume} {345}},\ \bibinfo {pages} {1484}
		(\bibinfo {year} {2014})}\BibitemShut {NoStop}%
	\bibitem [{\citenamefont {Yan}\ \emph {et~al.}(2013)\citenamefont {Yan},
		\citenamefont {Moses}, \citenamefont {Gadway}, \citenamefont {Covey},
		\citenamefont {Hazzard}, \citenamefont {Rey}, \citenamefont {Jin},\ and\
		\citenamefont {Ye}}]{Yan2013ood}%
	\BibitemOpen
	\bibfield  {author} {\bibinfo {author} {\bibfnamefont {B.}~\bibnamefont
			{Yan}}, \bibinfo {author} {\bibfnamefont {S.~A.}\ \bibnamefont {Moses}},
		\bibinfo {author} {\bibfnamefont {B.}~\bibnamefont {Gadway}}, \bibinfo
		{author} {\bibfnamefont {J.~P.}\ \bibnamefont {Covey}}, \bibinfo {author}
		{\bibfnamefont {K.~R.~A.}\ \bibnamefont {Hazzard}}, \bibinfo {author}
		{\bibfnamefont {A.~M.}\ \bibnamefont {Rey}}, \bibinfo {author} {\bibfnamefont
			{D.~S.}\ \bibnamefont {Jin}}, \ and\ \bibinfo {author} {\bibfnamefont
			{J.}~\bibnamefont {Ye}},\ }\href {http://dx.doi.org/10.1038/nature12483
		10.1038/nature12483
		http://www.nature.com/nature/journal/v501/n7468/abs/nature12483.html\#supplementary-information}
	{\bibfield  {journal} {\bibinfo  {journal} {Nature}\ }\textbf {\bibinfo
			{volume} {501}},\ \bibinfo {pages} {521} (\bibinfo {year}
		{2013})}\BibitemShut {NoStop}%
	\bibitem [{\citenamefont {de~Paz}\ \emph {et~al.}(2013)\citenamefont {de~Paz},
		\citenamefont {Sharma}, \citenamefont {Chotia}, \citenamefont {Mar\'{e}chal},
		\citenamefont {Huckans}, \citenamefont {Pedri}, \citenamefont {Santos},
		\citenamefont {Gorceix}, \citenamefont {Vernac},\ and\ \citenamefont
		{Laburthe-Tolra}}]{DePaz2013nqm}%
	\BibitemOpen
	\bibfield  {author} {\bibinfo {author} {\bibfnamefont {A.}~\bibnamefont
			{de~Paz}}, \bibinfo {author} {\bibfnamefont {A.}~\bibnamefont {Sharma}},
		\bibinfo {author} {\bibfnamefont {A.}~\bibnamefont {Chotia}}, \bibinfo
		{author} {\bibfnamefont {E.}~\bibnamefont {Mar\'{e}chal}}, \bibinfo {author}
		{\bibfnamefont {J.~H.}\ \bibnamefont {Huckans}}, \bibinfo {author}
		{\bibfnamefont {P.}~\bibnamefont {Pedri}}, \bibinfo {author} {\bibfnamefont
			{L.}~\bibnamefont {Santos}}, \bibinfo {author} {\bibfnamefont
			{O.}~\bibnamefont {Gorceix}}, \bibinfo {author} {\bibfnamefont
			{L.}~\bibnamefont {Vernac}}, \ and\ \bibinfo {author} {\bibfnamefont
			{B.}~\bibnamefont {Laburthe-Tolra}},\ }\href {\doibase
		10.1103/PhysRevLett.111.185305} {\bibfield  {journal} {\bibinfo  {journal}
			{Phys. Rev. Lett.}\ }\textbf {\bibinfo {volume} {111}},\ \bibinfo {pages}
		{185305} (\bibinfo {year} {2013})}\BibitemShut {NoStop}%
	\bibitem [{\citenamefont {Lu}\ \emph {et~al.}(2011)\citenamefont {Lu},
		\citenamefont {Burdick}, \citenamefont {Youn},\ and\ \citenamefont
		{Lev}}]{Lu2011sdb}%
	\BibitemOpen
	\bibfield  {author} {\bibinfo {author} {\bibfnamefont {M.}~\bibnamefont
			{Lu}}, \bibinfo {author} {\bibfnamefont {N.~Q.}\ \bibnamefont {Burdick}},
		\bibinfo {author} {\bibfnamefont {S.~H.}\ \bibnamefont {Youn}}, \ and\
		\bibinfo {author} {\bibfnamefont {B.~L.}\ \bibnamefont {Lev}},\ }\href
	{\doibase 10.1103/PhysRevLett.107.190401} {\bibfield  {journal} {\bibinfo
			{journal} {Phys. Rev. Lett.}\ }\textbf {\bibinfo {volume} {107}},\ \bibinfo
		{pages} {190401} (\bibinfo {year} {2011})}\BibitemShut {NoStop}%
	\bibitem [{\citenamefont {Sukachev}\ \emph {et~al.}(2010)\citenamefont
		{Sukachev}, \citenamefont {Sokolov}, \citenamefont {Chebakov}, \citenamefont
		{Akimov}, \citenamefont {Kanorsky}, \citenamefont {Kolachevsky},\ and\
		\citenamefont {Sorokin}}]{Sukachev2010mot}%
	\BibitemOpen
	\bibfield  {author} {\bibinfo {author} {\bibfnamefont {D.}~\bibnamefont
			{Sukachev}}, \bibinfo {author} {\bibfnamefont {A.}~\bibnamefont {Sokolov}},
		\bibinfo {author} {\bibfnamefont {K.}~\bibnamefont {Chebakov}}, \bibinfo
		{author} {\bibfnamefont {A.}~\bibnamefont {Akimov}}, \bibinfo {author}
		{\bibfnamefont {S.}~\bibnamefont {Kanorsky}}, \bibinfo {author}
		{\bibfnamefont {N.}~\bibnamefont {Kolachevsky}}, \ and\ \bibinfo {author}
		{\bibfnamefont {V.}~\bibnamefont {Sorokin}},\ }\href {\doibase
		10.1103/PhysRevA.82.011405} {\bibfield  {journal} {\bibinfo  {journal} {Phys.
				Rev. A}\ }\textbf {\bibinfo {volume} {82}},\ \bibinfo {pages} {011405}
		(\bibinfo {year} {2010})}\BibitemShut {NoStop}%
	\bibitem [{\citenamefont {Miao}\ \emph {et~al.}(2014)\citenamefont {Miao},
		\citenamefont {Hostetter}, \citenamefont {Stratis},\ and\ \citenamefont
		{Saffman}}]{Miao2014mot}%
	\BibitemOpen
	\bibfield  {author} {\bibinfo {author} {\bibfnamefont {J.}~\bibnamefont
			{Miao}}, \bibinfo {author} {\bibfnamefont {J.}~\bibnamefont {Hostetter}},
		\bibinfo {author} {\bibfnamefont {G.}~\bibnamefont {Stratis}}, \ and\
		\bibinfo {author} {\bibfnamefont {M.}~\bibnamefont {Saffman}},\ }\href
	{\doibase 10.1103/PhysRevA.89.041401} {\bibfield  {journal} {\bibinfo
			{journal} {Phys. Rev. A}\ }\textbf {\bibinfo {volume} {89}},\ \bibinfo
		{pages} {041401} (\bibinfo {year} {2014})}\BibitemShut {NoStop}%
	\bibitem [{\citenamefont {Petrov}\ \emph {et~al.}(2012)\citenamefont {Petrov},
		\citenamefont {Tiesinga},\ and\ \citenamefont {Kotochigova}}]{Petrov2012aif}%
	\BibitemOpen
	\bibfield  {author} {\bibinfo {author} {\bibfnamefont {A.}~\bibnamefont
			{Petrov}}, \bibinfo {author} {\bibfnamefont {E.}~\bibnamefont {Tiesinga}}, \
		and\ \bibinfo {author} {\bibfnamefont {S.}~\bibnamefont {Kotochigova}},\
	}\href {\doibase 10.1103/PhysRevLett.109.103002} {\bibfield  {journal}
	{\bibinfo  {journal} {Phys. Rev. Lett.}\ }\textbf {\bibinfo {volume} {109}},\
	\bibinfo {pages} {103002} (\bibinfo {year} {2012})}\BibitemShut {NoStop}%
\bibitem [{\citenamefont {Lepers}\ \emph {et~al.}(2014)\citenamefont {Lepers},
	\citenamefont {Wyart},\ and\ \citenamefont {Dulieu}}]{Lepers2014aot}%
\BibitemOpen
\bibfield  {author} {\bibinfo {author} {\bibfnamefont {M.}~\bibnamefont
		{Lepers}}, \bibinfo {author} {\bibfnamefont {J.-F.}\ \bibnamefont {Wyart}}, \
	and\ \bibinfo {author} {\bibfnamefont {O.}~\bibnamefont {Dulieu}},\ }\href
{\doibase 10.1103/PhysRevA.89.022505} {\bibfield  {journal} {\bibinfo
		{journal} {Phys. Rev. A}\ }\textbf {\bibinfo {volume} {89}},\ \bibinfo
	{pages} {022505} (\bibinfo {year} {2014})}\BibitemShut {NoStop}%
\bibitem [{\citenamefont {Frisch}\ \emph {et~al.}(2014)\citenamefont {Frisch},
	\citenamefont {Mark}, \citenamefont {Aikawa}, \citenamefont {Ferlaino},
	\citenamefont {Bohn}, \citenamefont {Makrides}, \citenamefont {Petrov},\ and\
	\citenamefont {Kotochigova}}]{Frisch2014qci}%
\BibitemOpen
\bibfield  {author} {\bibinfo {author} {\bibfnamefont {A.}~\bibnamefont
		{Frisch}}, \bibinfo {author} {\bibfnamefont {M.}~\bibnamefont {Mark}},
	\bibinfo {author} {\bibfnamefont {K.}~\bibnamefont {Aikawa}}, \bibinfo
	{author} {\bibfnamefont {F.}~\bibnamefont {Ferlaino}}, \bibinfo {author}
	{\bibfnamefont {J.~L.}\ \bibnamefont {Bohn}}, \bibinfo {author}
	{\bibfnamefont {C.}~\bibnamefont {Makrides}}, \bibinfo {author}
	{\bibfnamefont {A.}~\bibnamefont {Petrov}}, \ and\ \bibinfo {author}
	{\bibfnamefont {S.}~\bibnamefont {Kotochigova}},\ }\href
{http://dx.doi.org/10.1038/nature13137} {\bibfield  {journal} {\bibinfo
		{journal} {Nature}\ }\textbf {\bibinfo {volume} {507}},\ \bibinfo {pages}
	{475} (\bibinfo {year} {2014})}\BibitemShut {NoStop}%
\bibitem [{\citenamefont {Baumann}\ \emph {et~al.}(2014)\citenamefont
	{Baumann}, \citenamefont {Burdick}, \citenamefont {Lu},\ and\ \citenamefont
	{Lev}}]{Baumann2014ool}%
\BibitemOpen
\bibfield  {author} {\bibinfo {author} {\bibfnamefont {K.}~\bibnamefont
		{Baumann}}, \bibinfo {author} {\bibfnamefont {N.~Q.}\ \bibnamefont
		{Burdick}}, \bibinfo {author} {\bibfnamefont {M.}~\bibnamefont {Lu}}, \ and\
	\bibinfo {author} {\bibfnamefont {B.~L.}\ \bibnamefont {Lev}},\ }\href
{\doibase 10.1103/PhysRevA.89.020701} {\bibfield  {journal} {\bibinfo
		{journal} {Phys. Rev. A}\ }\textbf {\bibinfo {volume} {89}},\ \bibinfo
	{pages} {020701} (\bibinfo {year} {2014})}\BibitemShut {NoStop}%
\bibitem [{\citenamefont {Chin}\ \emph {et~al.}(2010)\citenamefont {Chin},
	\citenamefont {Grimm}, \citenamefont {Julienne},\ and\ \citenamefont
	{Tiesinga}}]{Chin2010fri}%
\BibitemOpen
\bibfield  {author} {\bibinfo {author} {\bibfnamefont {C.}~\bibnamefont
		{Chin}}, \bibinfo {author} {\bibfnamefont {R.}~\bibnamefont {Grimm}},
	\bibinfo {author} {\bibfnamefont {P.}~\bibnamefont {Julienne}}, \ and\
	\bibinfo {author} {\bibfnamefont {E.}~\bibnamefont {Tiesinga}},\ }\href
{\doibase 10.1103/RevModPhys.82.1225} {\bibfield  {journal} {\bibinfo
		{journal} {Rev. Mod. Phys.}\ }\textbf {\bibinfo {volume} {82}},\ \bibinfo
	{pages} {1225} (\bibinfo {year} {2010})}\BibitemShut {NoStop}%
\bibitem [{Sup()}]{SupplementMat}%
\BibitemOpen
\href@noop {} {}\bibinfo {note} {See the Supplemental Material for details
	regarding the creation of Er$_2$ molecules, the theoretical description of
	the collision formalism, and the adiabatic model.}\BibitemShut {Stop}%
\bibitem [{\citenamefont {Claussen}\ \emph {et~al.}(2003)\citenamefont
	{Claussen}, \citenamefont {Kokkelmans}, \citenamefont {Thompson},
	\citenamefont {Donley}, \citenamefont {Hodby},\ and\ \citenamefont
	{Wieman}}]{Claussen2003vhp}%
\BibitemOpen
\bibfield  {author} {\bibinfo {author} {\bibfnamefont {N.~R.}\ \bibnamefont
		{Claussen}}, \bibinfo {author} {\bibfnamefont {S.~J. J. M.~F.}\ \bibnamefont
		{Kokkelmans}}, \bibinfo {author} {\bibfnamefont {S.~T.}\ \bibnamefont
		{Thompson}}, \bibinfo {author} {\bibfnamefont {E.~A.}\ \bibnamefont
		{Donley}}, \bibinfo {author} {\bibfnamefont {E.}~\bibnamefont {Hodby}}, \
	and\ \bibinfo {author} {\bibfnamefont {C.~E.}\ \bibnamefont {Wieman}},\
}\href {\doibase 10.1103/PhysRevA.67.060701} {\bibfield  {journal} {\bibinfo
	{journal} {Phys. Rev. A}\ }\textbf {\bibinfo {volume} {67}},\ \bibinfo
{pages} {060701} (\bibinfo {year} {2003})}\BibitemShut {NoStop}%
\bibitem [{\citenamefont {Thompson}\ \emph {et~al.}(2005)\citenamefont
	{Thompson}, \citenamefont {Hodby},\ and\ \citenamefont
	{Wieman}}]{Thompson2005ump}%
\BibitemOpen
\bibfield  {author} {\bibinfo {author} {\bibfnamefont {S.~T.}\ \bibnamefont
		{Thompson}}, \bibinfo {author} {\bibfnamefont {E.}~\bibnamefont {Hodby}}, \
	and\ \bibinfo {author} {\bibfnamefont {C.~E.}\ \bibnamefont {Wieman}},\
}\href {\doibase 10.1103/PhysRevLett.95.190404} {\bibfield  {journal}
{\bibinfo  {journal} {Phys. Rev. Lett.}\ }\textbf {\bibinfo {volume} {95}},\
\bibinfo {eid} {190404} (\bibinfo {year} {2005})}\BibitemShut {NoStop}%
\bibitem [{\citenamefont {Mark}\ \emph
	{et~al.}(2007{\natexlab{a}})\citenamefont {Mark}, \citenamefont {Ferlaino},
	\citenamefont {Knoop}, \citenamefont {Danzl}, \citenamefont {Kraemer},
	\citenamefont {Chin}, \citenamefont {N\"{a}gerl},\ and\ \citenamefont
	{Grimm}}]{Mark2007sou}%
\BibitemOpen
\bibfield  {author} {\bibinfo {author} {\bibfnamefont {M.}~\bibnamefont
		{Mark}}, \bibinfo {author} {\bibfnamefont {F.}~\bibnamefont {Ferlaino}},
	\bibinfo {author} {\bibfnamefont {S.}~\bibnamefont {Knoop}}, \bibinfo
	{author} {\bibfnamefont {J.~G.}\ \bibnamefont {Danzl}}, \bibinfo {author}
	{\bibfnamefont {T.}~\bibnamefont {Kraemer}}, \bibinfo {author} {\bibfnamefont
		{C.}~\bibnamefont {Chin}}, \bibinfo {author} {\bibfnamefont {H.-C.}\
		\bibnamefont {N\"{a}gerl}}, \ and\ \bibinfo {author} {\bibfnamefont
		{R.}~\bibnamefont {Grimm}},\ }\href {\doibase 10.1103/PhysRevA.76.042514}
{\bibfield  {journal} {\bibinfo  {journal} {Phys. Rev. A}\ }\textbf {\bibinfo
		{volume} {76}},\ \bibinfo {pages} {042514} (\bibinfo {year}
	{2007}{\natexlab{a}})}\BibitemShut {NoStop}%
\bibitem [{Note1()}]{Note1}%
\BibitemOpen
\bibinfo {note} {For future experiments, it will be very interesting to
	access molecular levels parallel to the atomic threshold for which $\mu =2\mu
	_a$~\cite {Mark2007sou}.}\BibitemShut {Stop}%
\bibitem [{\citenamefont {Moerdijk}\ \emph {et~al.}(1995)\citenamefont
	{Moerdijk}, \citenamefont {Verhaar},\ and\ \citenamefont
	{Axelsson}}]{Moerdijk1995riu}%
\BibitemOpen
\bibfield  {author} {\bibinfo {author} {\bibfnamefont {A.~J.}\ \bibnamefont
		{Moerdijk}}, \bibinfo {author} {\bibfnamefont {B.~J.}\ \bibnamefont
		{Verhaar}}, \ and\ \bibinfo {author} {\bibfnamefont {A.}~\bibnamefont
		{Axelsson}},\ }\href {\doibase 10.1103/PhysRevA.51.4852} {\bibfield
	{journal} {\bibinfo  {journal} {Phys. Rev. A}\ }\textbf {\bibinfo {volume}
		{51}},\ \bibinfo {pages} {4852} (\bibinfo {year} {1995})}\BibitemShut
{NoStop}%
\bibitem [{\citenamefont {Stan}\ \emph {et~al.}(2004)\citenamefont {Stan},
	\citenamefont {Zwierlein}, \citenamefont {Schunck}, \citenamefont {Raupach},\
	and\ \citenamefont {Ketterle}}]{Stan2004oof}%
\BibitemOpen
\bibfield  {author} {\bibinfo {author} {\bibfnamefont {C.~A.}\ \bibnamefont
		{Stan}}, \bibinfo {author} {\bibfnamefont {M.~W.}\ \bibnamefont {Zwierlein}},
	\bibinfo {author} {\bibfnamefont {C.~H.}\ \bibnamefont {Schunck}}, \bibinfo
	{author} {\bibfnamefont {S.~M.~F.}\ \bibnamefont {Raupach}}, \ and\ \bibinfo
	{author} {\bibfnamefont {W.}~\bibnamefont {Ketterle}},\ }\href {\doibase
	10.1103/PhysRevLett.93.143001} {\bibfield  {journal} {\bibinfo  {journal}
		{Phys. Rev. Lett.}\ }\textbf {\bibinfo {volume} {93}},\ \bibinfo {pages}
	{143001} (\bibinfo {year} {2004})}\BibitemShut {NoStop}%
\bibitem [{\citenamefont {Wille}\ \emph {et~al.}(2008)\citenamefont {Wille},
	\citenamefont {Spiegelhalder}, \citenamefont {Kerner}, \citenamefont {Naik},
	\citenamefont {Trenkwalder}, \citenamefont {Hendl}, \citenamefont {Schreck},
	\citenamefont {Grimm}, \citenamefont {Tiecke}, \citenamefont {Walraven},
	\citenamefont {Kokkelmans}, \citenamefont {Tiesinga},\ and\ \citenamefont
	{Julienne}}]{Wille2008eau}%
\BibitemOpen
\bibfield  {author} {\bibinfo {author} {\bibfnamefont {E.}~\bibnamefont
		{Wille}}, \bibinfo {author} {\bibfnamefont {F.~M.}\ \bibnamefont
		{Spiegelhalder}}, \bibinfo {author} {\bibfnamefont {G.}~\bibnamefont
		{Kerner}}, \bibinfo {author} {\bibfnamefont {D.}~\bibnamefont {Naik}},
	\bibinfo {author} {\bibfnamefont {A.}~\bibnamefont {Trenkwalder}}, \bibinfo
	{author} {\bibfnamefont {G.}~\bibnamefont {Hendl}}, \bibinfo {author}
	{\bibfnamefont {F.}~\bibnamefont {Schreck}}, \bibinfo {author} {\bibfnamefont
		{R.}~\bibnamefont {Grimm}}, \bibinfo {author} {\bibfnamefont {T.~G.}\
		\bibnamefont {Tiecke}}, \bibinfo {author} {\bibfnamefont {J.~T.~M.}\
		\bibnamefont {Walraven}}, \bibinfo {author} {\bibfnamefont {S.~J. J. M.~F.}\
		\bibnamefont {Kokkelmans}}, \bibinfo {author} {\bibfnamefont
		{E.}~\bibnamefont {Tiesinga}}, \ and\ \bibinfo {author} {\bibfnamefont
		{P.~S.}\ \bibnamefont {Julienne}},\ }\href {\doibase
	10.1103/PhysRevLett.100.053201} {\bibfield  {journal} {\bibinfo  {journal}
		{Phys. Rev. Lett.}\ }\textbf {\bibinfo {volume} {100}},\ \bibinfo {eid}
	{053201} (\bibinfo {year} {2008})}\BibitemShut {NoStop}%
\bibitem [{\citenamefont {Tiecke}\ \emph {et~al.}(2010)\citenamefont {Tiecke},
	\citenamefont {Goosen}, \citenamefont {Walraven},\ and\ \citenamefont
	{Kokkelmans}}]{Tiecke2010abs}%
\BibitemOpen
\bibfield  {author} {\bibinfo {author} {\bibfnamefont {T.~G.}\ \bibnamefont
		{Tiecke}}, \bibinfo {author} {\bibfnamefont {M.~R.}\ \bibnamefont {Goosen}},
	\bibinfo {author} {\bibfnamefont {J.~T.~M.}\ \bibnamefont {Walraven}}, \ and\
	\bibinfo {author} {\bibfnamefont {S.~J. J. M.~F.}\ \bibnamefont
		{Kokkelmans}},\ }\href {\doibase 10.1103/PhysRevA.82.042712} {\bibfield
	{journal} {\bibinfo  {journal} {Phys. Rev. A}\ }\textbf {\bibinfo {volume}
		{82}},\ \bibinfo {pages} {042712} (\bibinfo {year} {2010})}\BibitemShut
{NoStop}%
\bibitem [{\citenamefont {Ni}\ \emph {et~al.}(2010)\citenamefont {Ni},
	\citenamefont {Ospelkaus}, \citenamefont {Wang}, \citenamefont
	{Qu\'em\'ener}, \citenamefont {Neyenhuis}, \citenamefont {de~Miranda},
	\citenamefont {Bohn}, \citenamefont {Ye},\ and\ \citenamefont
	{Jin}}]{Ni2010dco}%
\BibitemOpen
\bibfield  {author} {\bibinfo {author} {\bibfnamefont {K.~K.}\ \bibnamefont
		{Ni}}, \bibinfo {author} {\bibfnamefont {S.}~\bibnamefont {Ospelkaus}},
	\bibinfo {author} {\bibfnamefont {D.}~\bibnamefont {Wang}}, \bibinfo {author}
	{\bibfnamefont {G.}~\bibnamefont {Qu\'em\'ener}}, \bibinfo {author}
	{\bibfnamefont {B.}~\bibnamefont {Neyenhuis}}, \bibinfo {author}
	{\bibfnamefont {M.~H.~G.}\ \bibnamefont {de~Miranda}}, \bibinfo {author}
	{\bibfnamefont {J.~L.}\ \bibnamefont {Bohn}}, \bibinfo {author}
	{\bibfnamefont {J.}~\bibnamefont {Ye}}, \ and\ \bibinfo {author}
	{\bibfnamefont {D.~S.}\ \bibnamefont {Jin}},\ }\href {\doibase
	10.1038/nature08953} {\bibfield  {journal} {\bibinfo  {journal} {Nature}\
	}\textbf {\bibinfo {volume} {464}},\ \bibinfo {pages} {1324} (\bibinfo {year}
	{2010})}\BibitemShut {NoStop}%
\bibitem [{\citenamefont {Qu\'em\'ener}\ and\ \citenamefont
	{Bohn}(2010)}]{Quemener2010efs}%
\BibitemOpen
\bibfield  {author} {\bibinfo {author} {\bibfnamefont {G.}~\bibnamefont
		{Qu\'em\'ener}}\ and\ \bibinfo {author} {\bibfnamefont {J.~L.}\ \bibnamefont
		{Bohn}},\ }\href {\doibase 10.1103/PhysRevA.81.060701} {\bibfield  {journal}
	{\bibinfo  {journal} {Phys. Rev. A}\ }\textbf {\bibinfo {volume} {81}},\
	\bibinfo {pages} {060701} (\bibinfo {year} {2010})}\BibitemShut {NoStop}%
\bibitem [{\citenamefont {Micheli}\ \emph {et~al.}(2010)\citenamefont
	{Micheli}, \citenamefont {Idziaszek}, \citenamefont {Pupillo}, \citenamefont
	{Baranov}, \citenamefont {Zoller},\ and\ \citenamefont
	{Julienne}}]{Micheli2010urf}%
\BibitemOpen
\bibfield  {author} {\bibinfo {author} {\bibfnamefont {A.}~\bibnamefont
		{Micheli}}, \bibinfo {author} {\bibfnamefont {Z.}~\bibnamefont {Idziaszek}},
	\bibinfo {author} {\bibfnamefont {G.}~\bibnamefont {Pupillo}}, \bibinfo
	{author} {\bibfnamefont {M.~A.}\ \bibnamefont {Baranov}}, \bibinfo {author}
	{\bibfnamefont {P.}~\bibnamefont {Zoller}}, \ and\ \bibinfo {author}
	{\bibfnamefont {P.~S.}\ \bibnamefont {Julienne}},\ }\href {\doibase
	10.1103/PhysRevLett.105.073202} {\bibfield  {journal} {\bibinfo  {journal}
		{Phys. Rev. Lett.}\ }\textbf {\bibinfo {volume} {105}},\ \bibinfo {pages}
	{073202} (\bibinfo {year} {2010})}\BibitemShut {NoStop}%
\bibitem [{\citenamefont {de~Miranda}\ \emph {et~al.}(2011)\citenamefont
	{de~Miranda}, \citenamefont {Chotia}, \citenamefont {Neyenhuis},
	\citenamefont {Wang}, \citenamefont {Qu{\'e}m{\'e}ner}, \citenamefont
	{Ospelkaus}, \citenamefont {Bohn}, \citenamefont {Ye},\ and\ \citenamefont
	{Jin}}]{Miranda2011ctq}%
\BibitemOpen
\bibfield  {author} {\bibinfo {author} {\bibfnamefont {M.~H.~G.}\
		\bibnamefont {de~Miranda}}, \bibinfo {author} {\bibfnamefont
		{A.}~\bibnamefont {Chotia}}, \bibinfo {author} {\bibfnamefont
		{B.}~\bibnamefont {Neyenhuis}}, \bibinfo {author} {\bibfnamefont
		{D.}~\bibnamefont {Wang}}, \bibinfo {author} {\bibfnamefont {G.}~\bibnamefont
		{Qu{\'e}m{\'e}ner}}, \bibinfo {author} {\bibfnamefont {S.}~\bibnamefont
		{Ospelkaus}}, \bibinfo {author} {\bibfnamefont {J.~L.}\ \bibnamefont {Bohn}},
	\bibinfo {author} {\bibfnamefont {J.}~\bibnamefont {Ye}}, \ and\ \bibinfo
	{author} {\bibfnamefont {D.~S.}\ \bibnamefont {Jin}},\ }\href {\doibase
	10.1038/nphys1939} {\bibfield  {journal} {\bibinfo  {journal} {Nature Phys.}\
	}\textbf {\bibinfo {volume} {7}},\ \bibinfo {pages} {502} (\bibinfo {year}
	{2011})}\BibitemShut {NoStop}%
\bibitem [{Note2()}]{Note2}%
\BibitemOpen
\bibinfo {note} {The state $\mu _3$ is difficult to access because of its
	proximity with $\mu _4$.}\BibitemShut {Stop}%
\bibitem [{\citenamefont {Ferlaino}\ \emph {et~al.}(2010)\citenamefont
	{Ferlaino}, \citenamefont {Knoop}, \citenamefont {Berninger}, \citenamefont
	{Mark}, \citenamefont {N\"agerl},\ and\ \citenamefont
	{Grimm}}]{Ferlaino2010cou}%
\BibitemOpen
\bibfield  {author} {\bibinfo {author} {\bibfnamefont {F.}~\bibnamefont
		{Ferlaino}}, \bibinfo {author} {\bibfnamefont {S.}~\bibnamefont {Knoop}},
	\bibinfo {author} {\bibfnamefont {M.}~\bibnamefont {Berninger}}, \bibinfo
	{author} {\bibfnamefont {M.}~\bibnamefont {Mark}}, \bibinfo {author}
	{\bibfnamefont {H.-C.}\ \bibnamefont {N\"agerl}}, \ and\ \bibinfo {author}
	{\bibfnamefont {R.}~\bibnamefont {Grimm}},\ }\href {\doibase
	10.1134/S1054660X0917006X} {\bibfield  {journal} {\bibinfo  {journal} {Laser
			Phys.}\ }\textbf {\bibinfo {volume} {20}},\ \bibinfo {pages} {23} (\bibinfo
	{year} {2010})}\BibitemShut {NoStop}%
\bibitem [{\citenamefont {Qu\'em\'ener}\ \emph {et~al.}(2015)\citenamefont
	{Qu\'em\'ener}, \citenamefont {Lepers},\ and\ \citenamefont
	{Dulieu}}]{Quemener2015dou}%
\BibitemOpen
\bibfield  {author} {\bibinfo {author} {\bibfnamefont {G.}~\bibnamefont
		{Qu\'em\'ener}}, \bibinfo {author} {\bibfnamefont {M.}~\bibnamefont
		{Lepers}}, \ and\ \bibinfo {author} {\bibfnamefont {O.}~\bibnamefont
		{Dulieu}},\ }\href {\doibase 10.1103/PhysRevA.92.042706} {\bibfield
	{journal} {\bibinfo  {journal} {Phys. Rev. A}\ }\textbf {\bibinfo {volume}
		{92}},\ \bibinfo {pages} {042706} (\bibinfo {year} {2015})}\BibitemShut
{NoStop}%
\bibitem [{Note3()}]{Note3}%
\BibitemOpen
\bibinfo {note} {Note that this value is two orders of magnitude larger than
	that of $p$-wave dominated fermionic KRb molecules~\cite
	{Ni2010dco}.}\BibitemShut {Stop}%
\bibitem [{\citenamefont {Qu\'em\'ener}\ and\ \citenamefont
	{Bohn}(2011)}]{Quemener2011dou}%
\BibitemOpen
\bibfield  {author} {\bibinfo {author} {\bibfnamefont {G.}~\bibnamefont
		{Qu\'em\'ener}}\ and\ \bibinfo {author} {\bibfnamefont {J.~L.}\ \bibnamefont
		{Bohn}},\ }\href {\doibase 10.1103/PhysRevA.83.012705} {\bibfield  {journal}
	{\bibinfo  {journal} {Phys. Rev. A}\ }\textbf {\bibinfo {volume} {83}},\
	\bibinfo {pages} {012705} (\bibinfo {year} {2011})}\BibitemShut {NoStop}%
\bibitem [{\citenamefont {Qu\'em\'ener}\ \emph {et~al.}(2011)\citenamefont
	{Qu\'em\'ener}, \citenamefont {Bohn}, \citenamefont {Petrov},\ and\
	\citenamefont {Kotochigova}}]{Quemener2011uiu}%
\BibitemOpen
\bibfield  {author} {\bibinfo {author} {\bibfnamefont {G.}~\bibnamefont
		{Qu\'em\'ener}}, \bibinfo {author} {\bibfnamefont {J.~L.}\ \bibnamefont
		{Bohn}}, \bibinfo {author} {\bibfnamefont {A.}~\bibnamefont {Petrov}}, \ and\
	\bibinfo {author} {\bibfnamefont {S.}~\bibnamefont {Kotochigova}},\ }\href
{\doibase 10.1103/PhysRevA.84.062703} {\bibfield  {journal} {\bibinfo
		{journal} {Phys. Rev. A}\ }\textbf {\bibinfo {volume} {84}},\ \bibinfo
	{pages} {062703} (\bibinfo {year} {2011})}\BibitemShut {NoStop}%
\bibitem [{\citenamefont {Mark}\ \emph
	{et~al.}(2007{\natexlab{b}})\citenamefont {Mark}, \citenamefont {Kraemer},
	\citenamefont {Waldburger}, \citenamefont {Herbig}, \citenamefont {Chin},
	\citenamefont {N\"{a}gerl},\ and\ \citenamefont {Grimm}}]{Mark2007siw}%
\BibitemOpen
\bibfield  {author} {\bibinfo {author} {\bibfnamefont {M.}~\bibnamefont
		{Mark}}, \bibinfo {author} {\bibfnamefont {T.}~\bibnamefont {Kraemer}},
	\bibinfo {author} {\bibfnamefont {P.}~\bibnamefont {Waldburger}}, \bibinfo
	{author} {\bibfnamefont {J.}~\bibnamefont {Herbig}}, \bibinfo {author}
	{\bibfnamefont {C.}~\bibnamefont {Chin}}, \bibinfo {author} {\bibfnamefont
		{H.-C.}\ \bibnamefont {N\"{a}gerl}}, \ and\ \bibinfo {author} {\bibfnamefont
		{R.}~\bibnamefont {Grimm}},\ }\href
{http://link.aps.org/doi/10.1103/PhysRevLett.99.113201} {\bibfield  {journal}
	{\bibinfo  {journal} {Phys. Rev. Lett.}\ }\textbf {\bibinfo {volume} {99}},\
	\bibinfo {eid} {113201} (\bibinfo {year} {2007}{\natexlab{b}})}\BibitemShut
{NoStop}%
\bibitem [{\citenamefont {Lang}\ \emph {et~al.}(2008)\citenamefont {Lang},
	\citenamefont {{van der Straten}}, \citenamefont {Brandst\"atter},
	\citenamefont {Thalhammer}, \citenamefont {Winkler}, \citenamefont
	{Julienne}, \citenamefont {Grimm},\ and\ \citenamefont
	{Hecker~Denschlag}}]{Lang2008ctm}%
\BibitemOpen
\bibfield  {author} {\bibinfo {author} {\bibfnamefont {F.}~\bibnamefont
		{Lang}}, \bibinfo {author} {\bibfnamefont {P.}~\bibnamefont {{van der
				Straten}}}, \bibinfo {author} {\bibfnamefont {B.}~\bibnamefont
		{Brandst\"atter}}, \bibinfo {author} {\bibfnamefont {G.}~\bibnamefont
		{Thalhammer}}, \bibinfo {author} {\bibfnamefont {K.}~\bibnamefont {Winkler}},
	\bibinfo {author} {\bibfnamefont {P.~S.}\ \bibnamefont {Julienne}}, \bibinfo
	{author} {\bibfnamefont {R.}~\bibnamefont {Grimm}}, \ and\ \bibinfo {author}
	{\bibfnamefont {J.}~\bibnamefont {Hecker~Denschlag}},\ }\href {\doibase
	10.1038/nphys838} {\bibfield  {journal} {\bibinfo  {journal} {Nature Phys.}\
	}\textbf {\bibinfo {volume} {4}},\ \bibinfo {pages} {223} (\bibinfo {year}
	{2008})}\BibitemShut {NoStop}%
\bibitem [{\citenamefont {Lepers}\ \emph {et~al.}(2015)\citenamefont {Lepers},
	\citenamefont {Qu\'em\'ener}, \citenamefont {Luc-Koenig},\ and\ \citenamefont
	{Dulieu}}]{Quemener2015fbl}%
\BibitemOpen
\bibfield  {author} {\bibinfo {author} {\bibfnamefont {M.}~\bibnamefont
		{Lepers}}, \bibinfo {author} {\bibfnamefont {G.}~\bibnamefont
		{Qu\'em\'ener}}, \bibinfo {author} {\bibfnamefont {E.}~\bibnamefont
		{Luc-Koenig}}, \ and\ \bibinfo {author} {\bibfnamefont {O.}~\bibnamefont
		{Dulieu}},\ }\href@noop {} {\bibfield  {journal} {\bibinfo  {journal}
		{arXiv:1508.06066 [J.Phys.B (to be published)]}\ } (\bibinfo {year}
	{2015})}\BibitemShut {NoStop}%
\bibitem [{\citenamefont {Greiner}\ \emph {et~al.}(2003)\citenamefont
	{Greiner}, \citenamefont {Regal},\ and\ \citenamefont
	{Jin}}]{Greiner2003eoa}%
\BibitemOpen
\bibfield  {author} {\bibinfo {author} {\bibfnamefont {M.}~\bibnamefont
		{Greiner}}, \bibinfo {author} {\bibfnamefont {C.~A.}\ \bibnamefont {Regal}},
	\ and\ \bibinfo {author} {\bibfnamefont {D.~S.}\ \bibnamefont {Jin}},\ }\href
{\doibase 10.1038/nature02199} {\bibfield  {journal} {\bibinfo  {journal}
		{Nature}\ }\textbf {\bibinfo {volume} {426}},\ \bibinfo {pages} {537}
	(\bibinfo {year} {2003})}\BibitemShut {NoStop}%
\bibitem [{\citenamefont {Jochim}\ \emph {et~al.}(2003)\citenamefont {Jochim},
	\citenamefont {Bartenstein}, \citenamefont {Altmeyer}, \citenamefont {Hendl},
	\citenamefont {Riedl}, \citenamefont {Chin}, \citenamefont {{Hecker
			Denschlag}},\ and\ \citenamefont {Grimm}}]{Jochim2003bec}%
\BibitemOpen
\bibfield  {author} {\bibinfo {author} {\bibfnamefont {S.}~\bibnamefont
		{Jochim}}, \bibinfo {author} {\bibfnamefont {M.}~\bibnamefont {Bartenstein}},
	\bibinfo {author} {\bibfnamefont {A.}~\bibnamefont {Altmeyer}}, \bibinfo
	{author} {\bibfnamefont {G.}~\bibnamefont {Hendl}}, \bibinfo {author}
	{\bibfnamefont {S.}~\bibnamefont {Riedl}}, \bibinfo {author} {\bibfnamefont
		{C.}~\bibnamefont {Chin}}, \bibinfo {author} {\bibfnamefont {J.}~\bibnamefont
		{{Hecker Denschlag}}}, \ and\ \bibinfo {author} {\bibfnamefont
		{R.}~\bibnamefont {Grimm}},\ }\href {\doibase 10.1126/science.1093280}
{\bibfield  {journal} {\bibinfo  {journal} {Science}\ }\textbf {\bibinfo
		{volume} {302}},\ \bibinfo {pages} {2101} (\bibinfo {year}
	{2003})}\BibitemShut {NoStop}%
\bibitem [{\citenamefont {Frisch}\ \emph {et~al.}(2012)\citenamefont {Frisch},
	\citenamefont {Aikawa}, \citenamefont {Mark}, \citenamefont {Rietzler},
	\citenamefont {Schindler}, \citenamefont {Zupanic}, \citenamefont {Grimm},\
	and\ \citenamefont {Ferlaino}}]{Frisch2012nlm}%
\BibitemOpen
\bibfield  {author} {\bibinfo {author} {\bibfnamefont {A.}~\bibnamefont
		{Frisch}}, \bibinfo {author} {\bibfnamefont {K.}~\bibnamefont {Aikawa}},
	\bibinfo {author} {\bibfnamefont {M.}~\bibnamefont {Mark}}, \bibinfo {author}
	{\bibfnamefont {A.}~\bibnamefont {Rietzler}}, \bibinfo {author}
	{\bibfnamefont {J.}~\bibnamefont {Schindler}}, \bibinfo {author}
	{\bibfnamefont {E.}~\bibnamefont {Zupanic}}, \bibinfo {author} {\bibfnamefont
		{R.}~\bibnamefont {Grimm}}, \ and\ \bibinfo {author} {\bibfnamefont
		{F.}~\bibnamefont {Ferlaino}},\ }\href {\doibase 10.1103/PhysRevA.85.051401}
{\bibfield  {journal} {\bibinfo  {journal} {Phys. Rev. A}\ }\textbf {\bibinfo
		{volume} {85}},\ \bibinfo {pages} {051401(R)} (\bibinfo {year}
	{2012})}\BibitemShut {NoStop}%
\bibitem [{\citenamefont {Idziaszek}\ and\ \citenamefont
	{Julienne}(2010)}]{Idziaszek2010urc}%
\BibitemOpen
\bibfield  {author} {\bibinfo {author} {\bibfnamefont {Z.}~\bibnamefont
		{Idziaszek}}\ and\ \bibinfo {author} {\bibfnamefont {P.~S.}\ \bibnamefont
		{Julienne}},\ }\href {\doibase 10.1103/PhysRevLett.104.113202} {\bibfield
	{journal} {\bibinfo  {journal} {Phys. Rev. Lett.}\ }\textbf {\bibinfo
		{volume} {104}},\ \bibinfo {pages} {113202} (\bibinfo {year}
	{2010})}\BibitemShut {NoStop}%
\bibitem [{\citenamefont {Kotochigova}\ and\ \citenamefont
	{Petrov}(2011)}]{Kotochigova2011ait}%
\BibitemOpen
\bibfield  {author} {\bibinfo {author} {\bibfnamefont {S.}~\bibnamefont
		{Kotochigova}}\ and\ \bibinfo {author} {\bibfnamefont {A.}~\bibnamefont
		{Petrov}},\ }\href {\doibase 10.1039/C1CP21175G} {\bibfield  {journal}
	{\bibinfo  {journal} {Phys. Chem. Chem. Phys.}\ }\textbf {\bibinfo {volume}
		{13}},\ \bibinfo {pages} {19165} (\bibinfo {year} {2011})}\BibitemShut
{NoStop}%
\end{thebibliography}

%

\clearpage

\setcounter{figure}{0}
\renewcommand{\thefigure}{S\arabic{figure}}

\section{Supplemental Material}

\subsection{Creation of Er$_2$ in 3D and Q2D}

We create Feshbach molecules using standard techniques of magneto-association across a Feshbach resonance. As demonstrated in Refs.\,\cite{Aikawa2012bec,Frisch2014qci}, Er features an enormous number of Feshbach resonances. Here, we focus on the resonances observed below $\unit[3]{G}$. In particular, we first create an ultracold atomic sample of about ${3\times10^5}$ $^{168}$Er atoms at a temperature of $T\approx \unit[150]{nK}$, which is just above the onset of Bose condensation, see Ref.\,\cite{Aikawa2012bec}. The atoms are confined into a three-dimensional (3D) crossed optical dipole trap with frequencies $\nu_x=\unit[51.5(2)]{Hz}$, $\nu_y=\unit[13.2(3)]{Hz}$, and $\nu_z=\unit[207(1)]{Hz}$. We choose magnetic fields of $\unit[1.4]{G}$, $\unit[2.3]{G}$, and $\unit[2.8]{G}$ for the molecular states $\mu_1$, $\mu_2$, and $\mu_4$, respectively. We then magneto-associate molecules by ramping the magnetic field $\unit[150]{mG}$ below the Feshbach resonance. The typical ramp speed is $\unit[90]{mG/ms}$. After the molecule association, we remove all the residual atoms from the optical dipole trap by applying a short laser pulse. The pulse is on resonance with the strong atomic transition at $\unit[401]{nm}$~\cite{Frisch2012nlm} and has a duration of $\unit[1]{\mu s}$ with an intensity of $\sim \unit[40]{mW/cm^2}$.

To realize a Q2D geometry, we superimpose a one dimensional optical lattice beam to the system after finishing evaporation in the 3D trap. The lattice is realized from a retro-reflected laser beam at $\unit[1064]{nm}$, propagating along the vertical direction. The beam has a waist of $\unit[250]{\mu m}$ and a typical power of $\unit[8]{W}$. As a result, the particles are confined into an array of Q2D pancakes with frequencies $\nu_r=\unit[33.0(3)]{Hz}$ in the radial direction and $\nu_z = \unit[31.2(1)]{kHz}$ in the tightly confining axial direction. We first load the lattice from the atomic sample and we then magneto-associate Er$_2$ in the lattice. The molecule conversion efficiency in the Q2D geometry is $\lesssim \unit[5]{\%}$, which is below the one observed in the 3D trap. With this scheme, we produce about $1.1 \times 10^4$ molecules at a temperature of $\unit[400]{nK}$, corresponding to a density of $\unit[3.8 \times 10^{7}]{cm^{-2}}$. The molecules fill about $35$ lattice layers.

We control the molecular dipole orientation by changing the orientation of the magnetic field. The orientation is quantified in term of the angle $\theta$, which defines the angle between the quantization axis, set by the magnetic field orientation, and the $z$-axis of the lattice trap. We prepare the molecular samples at either $\theta = 0^{\circ}$ or $90^{\circ}$, correspondingly side-by-side (repulsive) or head-to-tail (attractive) dipolar collisions.
The magnetic field is rotated by using three pairs of independently-controlled magnetic-field coils. We pay particular attention that when changing the orientation of the magnetic field we keep its magnitude constant. We check this by performing radio-frequency spectroscopy between Zeeman sub-levels for different angles of rotation. We typically rotate the field within $\sim \unit[6]{ms}$.

For all our loss-rate measurements, we jump to a magnetic field of about $\unit[200]{mG}$ after molecule association. At this field, $E_b$ is of the order of few $h\times\unit[1]{MHz}$. We choose to perform our measurement at this magnetic-field value because around $\unit[200]{mG}$ there are no Feshbach resonances and the molecular spectrum might be less dense. Using a Stern-Gerlach technique~\cite{Mark2007sou}, we measure $\mu$ at $\unit[200]{mG}$ for all the three target molecular states. We find a slight shift of $\mu$ in comparison with the values from Table~II of a few percent to $10.9(5)\,\mu_{B}$ for $\mu_1$, $11.7(3)\,\mu_{B}$ for $\mu_2$, and $8.7(6)\,\mu_{B}$ for $\mu_4$.

The given uncertainties for the measured loss rates in Table~II are composed of a statistical error with one standard deviation derived from fitting a two-body rate equation to the measured data, and a systematic uncertainty coming from number density calibration. Due to the distribution of molecules across many lattice layers this is by far the greatest uncertainty in the Q2D geometry. The average 2D density and its uncertainty was calculated using a number-weighted average over occupied lattice layers similar to Ref.\,\cite{Miranda2011ctq}. When calculating the loss rate ratio $\beta_\perp / \beta_\|$, the systematic uncertainty in the density can be neglected as it is highly correlated for the measurement of $\beta_\perp$ and $\beta_\|$.

\subsection{Collision Formalism}

We briefly describe the theoretical formalism used in this article to determine the collisional properties of Er$_2$ molecules in free space (3D collisions) and in an one-dimensional optical lattice (Q2D collisions), in an arbitrary magnetic field $\vec{B}$. More details can be found in Ref.\,\cite{Quemener2011dou,Quemener2015dou}.

We use a time-independent quantum formalism based on spherical coordinates $\vec{r}=(r,\theta_r,\phi_r)$ describing the relative motion of two Er$_2$ molecules. The quantization axis $\hat{z}$ is chosen to be the confinement axis of the optical lattice. A spherical harmonic basis set, summed over different partial waves $\ell$ with projections $m_\ell$ on the quantization axis, is used to expand the total colliding wave function. The one dimensional optical lattice is supposed to be deep enough to consider the collision taking place in an individual pancake. One pancake is represented as an harmonic trap for the relative motion of reduced mass $m_\text{red}$
\begin{eqnarray}
	V_{\mathrm{ho}} = \frac{1}{2} m_\mathrm{red} \, \omega^2 z^2 
\end{eqnarray}
with $\omega = 2 \pi \nu$ and $\nu = \unit[31.2]{kHz}$. The 3D collisions are recovered by setting $\nu = 0$. We consider molecules in the ground state of the harmonic oscillator. A given state of an Er$_2$ Feshbach molecule is described by a rather complicated linear combination of atomic states which cannot be precisely calculated as mentioned in the next section of this Supplemental Material. Therefore we consider that the molecule has a magnetic moment of magnitude $\mu$ aligned along the magnetic field which makes an angle $\theta$ with the confinement axis. The interaction between two molecules is provided by the magnetic dipole-dipole interaction
\begin{eqnarray}
	V_{\mathrm{dd}} = \frac{\mu^2 \,(1-3\cos^2(\theta_r-\theta))}{(4\pi/\mu_0)\, r^3} .
\end{eqnarray}
We also used an isotropic Er$_2$ + Er$_2$ van der Waals interaction given by
\begin{eqnarray}
	V_{\mathrm{vdW}} = - \frac{C_6}{r^6}
\end{eqnarray}
with $C_6 = 4 \times 1760 = \unit[7040]{a.u.}$ which amounts to four times the value of an isotropic atom-atom coefficient of $\unit[1760]{a.u.}$ from the theoretical work of Ref.\,\cite{Lepers2014aot}. 
Note that an alternative value of $\unit[1723]{a.u.}$ based on observed transitions was obtained in Ref.\,\cite{Frisch2014qci}. The Schr\"odinger equation is solved for each radial intermolecular separations $r$ using a log-derivative propagation method. Matching the colliding wavefunction and its derivative with appropriate two-dimensional asymptotic boundary conditions at long-range\,\cite{Quemener2011dou} provides the cross section and the rate coefficient as a function of the collision energy for any arbitrary configurations of magnetic fields and confinements. Averaging the cross sections over a 3D and 2D dimensional Maxwell-Boltzmann distribution provides the corresponding thermalized rate coefficients $\beta(T)$ for a given temperature.

At short range, we assume that the molecules undergo a full loss mechanism process with a unit probability (it can be either an inelastic or a possible reactive process). This assumption, which corresponds to the so-called universal regime in ultracold collisions, considers that the physics is independent of the initial short-range scattering phase-shift~\cite{Idziaszek2010urc} of the full potential energy surfaces of Er$_4$. This is what it is usually assumed for theory
as nothing is known about this potential energy surface at short range. Then, if the magnitude of the rates differs between experiment and theory, one can learn that an experimental system deviates from this universal regime and short-range effects play a role.

To circumvent this, it is more convenient to compute the ratio of the theoretical rates of two different magnetic field orientations since we will start with the same short-range physics condition for both orientations, and compare it with the corresponding experimental ratio. An analysis based on the universal behavior of dipolar collisions in confinement of Ref.\,\cite{Quemener2011uiu} using a Quantum Threshold model leads to the following formula for the ratio $\beta_\perp(T) / \beta_\|(T)$. For bosons, using Eq.\,30 of Ref.\,\cite{Quemener2011uiu} to describe $\beta_\perp$ (dipole dominated) and Eq.\,32 of the same reference for $\beta_\|$ (confinement dominated) we find
\begin{eqnarray}
	\frac{\beta_\perp(T)}{\beta_\|(T)}\bigg|_{\mathrm{bos}} \sim \left( \frac{a_{\mathrm{dB}}}{a_{\mathrm{ho}}}\right)^4 \ \frac{a_d}{a_{\mathrm{ho}}} \ e^{2(a_d/a_{\mathrm{ho}})^{2/5}}
\end{eqnarray}
when $a_d, a_{\mathrm{ho}} > a_{\mathrm{vdW}} $ for a fixed value of $a_\mathrm{dB}/a_{\mathrm{ho}}$ where $a_{\mathrm{dB}}$ is the thermal de Broglie wavelength.
For fermions, using Eq.\,16 of Ref.\,\cite{Quemener2011uiu} to describe $\beta_\perp$ (dipole dominated) and Eq.\,14 of the same reference for $\beta_\|$ (van der Waals dominated), along with Eq.\,27, we find
\begin{eqnarray}
	\frac{\beta_\perp}{\beta_\|}\bigg|_{\mathrm{fer}} \sim \left( \frac{a_d}{a_{\mathrm{vdW}}}\right)^3
\end{eqnarray}
when $a_d, a_{\mathrm{vdW}} < a_{\mathrm{ho}} $. These formulas suggest to plot the ratio as a function of $a_d/a_{\mathrm{ho}}$ for bosons for a fixed ratio $a_{\mathrm{dB}}/a_{\mathrm{ho}} = 2 \pi \sqrt{\nu/k_BT}$ and as a function of $a_d/a_{\mathrm{vdW}}$ for fermions, as it has been done in Fig.\,4 for the magnetic dipolar molecules of Er$_2$ and the electric polar molecules of KRb.

\subsection{Adiabatic Model}

In Ref.\,\cite{Frisch2014qci} we presented the theoretical bosonic-erbium Feshbach spectra derived from coupled-channels calculations. We concluded there that such first-principle evaluations can not quantitatively capture the complex scattering behavior of Er. In fact with the current computing capabilities, the calculations can not be converged with respect to the number of basis states required to explain the experimental Feshbach-resonance density. For this reason, we developed a novel approach based on adiabatic potentials (adiabats) $U_n(R;B)$.

Our adiabatic model starts from the Hamiltonian $H= -(\hbar^2/2m_r)d^2/dR^2 +V({\vec R})$. The first term is the radial kinetic-energy operator with $\vec R$ describing the orientation and the separation between the two atomic dipoles, and $m_r$ is the reduced mass. The second term of the Hamiltonian is the potential operator $V({\vec R})$, which describes the Zeeman and interatomic interactions. It reads $V({\vec R}) = \hbar^2 \vec{\ell}^2/(2m_r R^2) + H_Z +W^{\rm elec}(\vec R)$ and incorporates the rotational energy operator with molecular orbital angular momentum $\vec\ell$, the Zeeman interaction of two atoms $H_Z$, and the electronic potential operator $W^{\rm elec}(\vec R)$ between the particles. Our model assumes that the relative vibrational motion of two Er atoms is slow compared to the timescales of the rotational, Zeeman, and ``electronic'' atom-atom interactions.

The Zeeman interaction is $H_Z=g\mu_B (j_{1z}+j_{2z}) B$. Here, $g=1.16$ is the Er g-factor, a magnetic field $B$ is aligned along the $\hat z$ direction, and $j_{iz}$ is the $z$ component of the angular momentum operator $\vec\jmath_i$ of atom $i=1,2$. The electronic potential operator $W^{\rm elec}(\vec R)$, described in Refs.\,\cite{Kotochigova2011ait,Petrov2012aif,Frisch2014qci}, is anisotropic, as it depends on the orientation of $\vec R$. At large separation $R$, $W^{\rm elec}(\vec R)$ is given by the magnetic dipole-dipole interaction plus both the isotropic and anisotropic contribution of the van der Waals interaction. For $R\to\infty$ the interaction $W^{\rm elec}(\vec R)\to0$.

The Hamiltonian is evaluated in the basis $|i\rangle=|(j_1j_2)JM\rangle Y_{\ell m_\ell}(\hat R)$, where $\vec J=\vec\jmath_1+\vec\jmath_2$ and $Y_{\ell m_\ell}(\hat R)$ is a spherical harmonic. It conserves $m_\ell + M$ and parity $p=(-1)^{\ell}$. In addition, for bosonic isotopes $(-1)^{\ell+J}=1$. We focus on ultracold collisions between atomic states $|j_1m_{1}\rangle=|j_2m_{2}\rangle=|6,-6\rangle$ and, therefore, only include basis functions satisfying $m_{\ell} + M =-12$. We limit the included partial waves to even $\ell\le 6$ and thus to states with even $J$, as the ``adiabatic'' magnetic moments of the resonances quickly converge with the included number of partial waves (In our calculation there is one $s$-wave channel, four $d$-wave channels, nine $g$-wave channels, and 16 $i$-wave channels.).

The adiabats $U_n(R;B)$ with $n=1,2,\dots$ are eigenvalues of the operator $V({\vec R})$ at a given field strength $B$. Their eigenfunctions are $|n;R\rangle=\sum_{i} c_{n,i}(R)|i\rangle$ with $R$-dependent coefficients $c_{n,i}(R)$. Note that we neglect the coupling between $U_n(R;B)$ due to the radial part of kinetic-energy operator.

\begin{figure}
	\includegraphics[width=1.0\columnwidth]{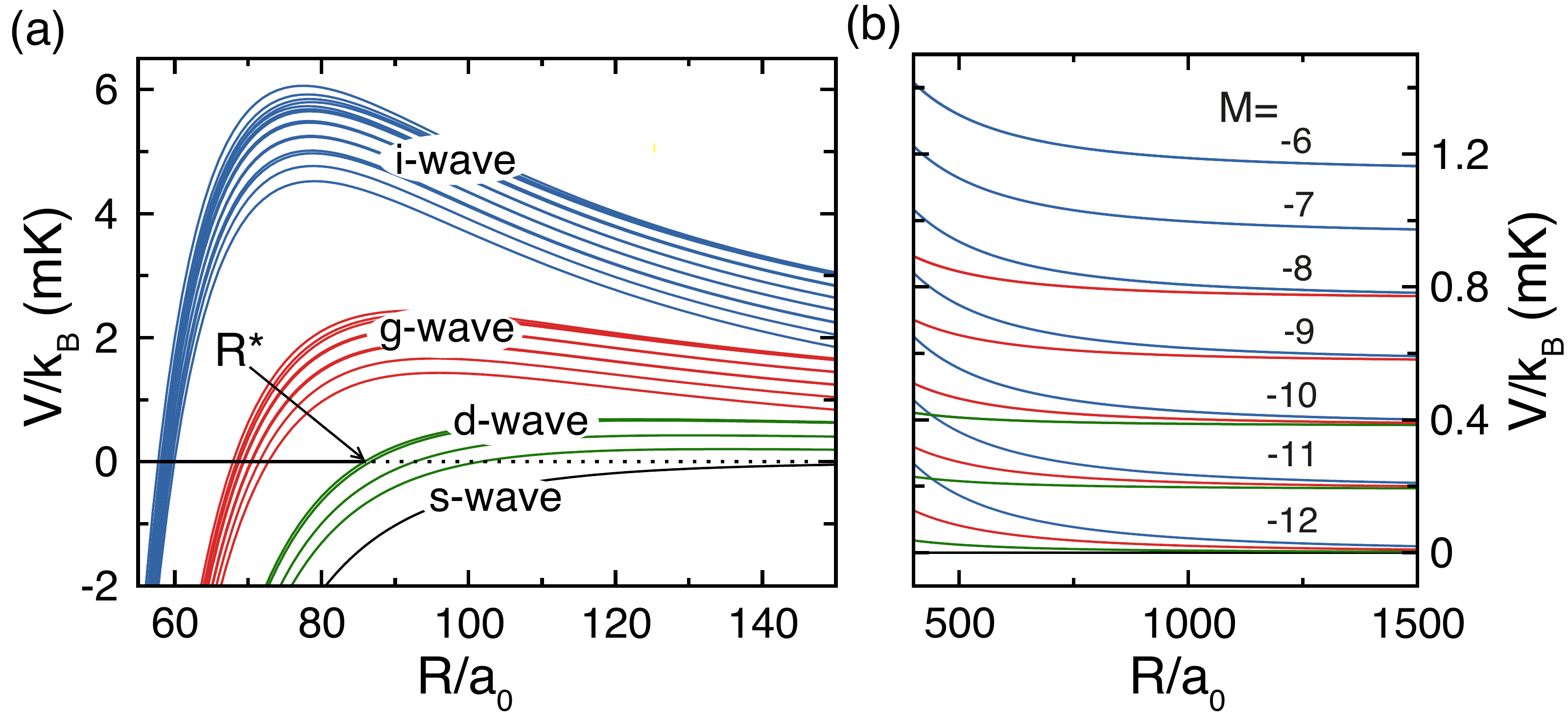}
	\caption{Adiabatic interaction potentials of two Er atoms at medium- (a) and long- (b) interatomic separation $R$. The calculation is performed at $B$ = 2.44 G with $m_{\ell} + M = -12$ and includes only states with even $\ell \le 6$. The zero of energy is at the dissociation limit of two $|j,m\rangle = |6,-6\rangle$ atoms. Black, green, red, and blue curves indicate the dominant $\ell$-wave character. The horizontal black line indicates a $d$-wave Feshbach molecule with an outer turning point $R = R^*$ resonant with the $s$-wave entrance channel. Panel (b) also shows the $M$ projection for each of the Zeeman dissociation limit.}
	\label{fig:adiabat}
\end{figure}

Figure~\ref{fig:adiabat} shows the adiabats at $B=\unit[2.44]{G}$. The scattering starts from the $s$-wave entrance channel correlating to the energetically lowest adiabat. All other potentials either have a centrifugal barrier and dissociate to two atoms with $M=-12$, or dissociate to closed-channel Zeeman sublevels with $M> -12$. We distinguish four groups of potentials, each associated with a dominant partial wave $\ell$. Within a group, the potentials are split by the Zeeman energy and the magnetic DDI and dissociate at different atomic thresholds. For each potential $U_n(R;B)$ we can further assign the dominant $J$ and $M$, where $\vec J=\vec\jmath_1+\vec\jmath_2$ is the sum of electronic angular momenta of two atoms and $M$ is the projection of $J$ on the internuclear axis. The figure also shows an example of predominantly $d$-wave Feshbach molecules with an outer classical turning point $R^*$. Its ``adiabatic'' molecular magnetic moment is to good approximation given by $\mu^{\rm calc}\approx-dU_n(R^*;B)/dB$, where we further use that most of the vibrational wavefunction is localized around $R^*$. Interestingly, we observe that the $\mu^{\rm calc}$ value quickly converges with the number of included $\ell$ (even $\ell\le6$ is sufficient) and that it strongly depends on the DDI but only weakly on the vdW dispersion potential. In fact, at $R^*$ the DDI dominates over the anisotropic part of the dispersion potential.

The adiabatic magnetic moment of a resonance is given by $\mu_{nv}^{\rm adiab} \equiv -dE_{nv}(B)/dB\approx-dU_n(R^*;B)/dB$, where we realize that to good approximation most of the adiabatic vibrational wavefunction is localized around the outer classical turning point. We further note that $dU_n(R^*;B)/dB = \langle n;R^*| dH_Z/dB| n;R^*\rangle$ from the Hellmann-Feyman theorem and, hence, $\mu_{nv}^{\rm adiab}=-g\mu_B \sum_{i}M_i|c_{i}(R^*)|^2$, where $M_i$ is the total atomic projection quantum number of state $|i\rangle$. We assign a resonance by the quantum numbers of the basis state $|i\rangle$ for which $|c_{i}(R^*)|^2$ is largest and note that the absolute value of the magnetic moment of a resonance is always smaller that $12g\mu_B\approx 14 \mu_B$.

We further assume that the non-adiabatic coupling between the adiabatic potentials is significantly smaller than their spacings for $R<100 a_0$. Then a weakly bound level of adiabatic potential $n$ can lead to a Feshbach resonance when its energy $E_{nv}(B)$ coincides with the entrance channel energy. The outer turning point $R^*$ of this level satisfies $U_n(R;B) = 0$. The resonance acquires a width due to non-adiabatic coupling to the entrance channel.

Finally, we determine the approximate quantum numbers of experimentally-observed resonances with $B_{\rm res}< 3$ G, listed in Table~I, based on a comparison of the experimental magnetic moment with those predicted by the adiabatic model at the same resonant field. We find that for these resonances there exist adiabats with a magnetic moment that agrees within 1\% uncertainty with the experimental values. A study of the largest coefficients $c_{n,i}(R)$
at $R=R^*$ then enables us to assign the dominant quantum states shown in Table~I.

Figure~2 (main text) shows the magnetic-field dependence of the adiabatic magnetic moment at the entrance channel energy for each of the adiabatic potentials $U_n(R;B)$. We see that for $B > 10$ G the magnetic moment values equal integer multiples of $g\mu_B$ corresponding to those of the atomic limits. For smaller field strengths the adiabatic magnetic moments show mixing of the Zeeman sublevels. Here, the magnetic moment value depends on the magnetic dipole-dipole interaction but only weakly on the strength and anisotropy of the dispersion potential. The figure also shows our experimentally studied Feshbach resonance locations as well as their magnetic moments $\mu^{\rm exp}$; see Table~I.

\end{document}